\documentclass[
    reprint,amsmath,amssymb,aps,nofootinbib,preprintnumbers,prl]{revtex4-2}

\usepackage{graphicx,adjustbox} 
\usepackage{dcolumn}
\usepackage{bm}
\usepackage[makeroom]{cancel}
\usepackage{url}

\usepackage{breakurl}
\usepackage{graphicx}
\usepackage{xcolor} 
\usepackage{esint} 
\usepackage{xr-hyper}
\usepackage{tikz-feynman}
\usepackage{caption,subcaption}
\DeclareMathOperator{\arccosh}{arccosh}

\usepackage[breaklinks]{hyperref}
\hypersetup{
	colorlinks=true,
	linktoc=page,
	citecolor=americanrose,
	linkcolor=cadmiumgreen,
	urlcolor=blue} 
\def\appendixautorefname~#1\null{Appendix\,#1\null}
\def\sectionautorefname~#1\null{Section\,#1\null}
\def\subsectionautorefname~#1\null{Section\,#1\null}
\def\figureautorefname~#1\null{Figure\,#1\null}
\def\equationautorefname~#1\null{Equation\,(#1)\null}

\definecolor{americanrose}{rgb}{1.0, 0.01, 0.24}
\definecolor{cadmiumgreen}{rgb}{0.0, 0.42, 0.24}

\usepackage{tikz}
\usepackage{tikz-feynman}
\usetikzlibrary{positioning}
\tikzfeynmanset{compat=1.1.0}
\definecolor{nodefill}{RGB}{232,232,236} 

\newcommand{\HalfLen}{1.5}   
\newcommand{\Ysep}{0.58}     
\newcommand{\BlobR}{7mm}     
\tikzset{
  kernelline/.style={line width=0.9pt, line cap=round, line join=round},
  blobstyle/.style={circle, draw=black, line width=0.9pt, fill=gray!15, minimum size=2*\BlobR},
  gluon/.style={decorate, decoration={coil, aspect=0, segment length=1.6mm, amplitude=0.75mm}, line width=0.9pt}
}

\usepackage[colorinlistoftodos]{todonotes}
\setuptodonotes{color=cyan}
\presetkeys{todonotes}{inline}{}
\reversemarginpar 
\makeatletter\g@addto@macro\bfseries{\boldmath}\makeatother

\newcommand{\DB}{\scalebox{0.6}{\ensuremath{\mathrm{DB}}}}
\newcommand{\PB}{\scalebox{0.6}{\ensuremath{\mathrm{PB}}}}

\newcommand{\mathcalMtree}{\mathcal{M}^{(0)}_{4,\raisebox{0.8ex}{\scalebox{0.16}{
\begin{tikzpicture}[baseline={(0,-0.1)},line width=1pt]
  \begin{feynman}
    \diagram [horizontal=a to b] {
      i1 -- [solid] a -- [solid] f1,
      i2 -- [solid] b -- [solid] f2,
      a -- [boson] b,
    };
  \end{feynman}
\end{tikzpicture}
}}} }

\begin{document}

\preprint{QMUL-PH-25-20}
\title{Dirac brackets for classical radiative observables} 

\author{Francesco Alessio}\email{Francesco.Alessio@lnf.infn.it}
\affiliation{INFN, Laboratori Nazionali di Frascati, 00044 Frascati (RM), Italy }%
\author{Riccardo Gonzo}\email{rgonzo@ed.ac.uk} 
\affiliation{Higgs Centre for Theoretical Physics, School of Physics and Astronomy, 
University of Edinburgh, EH9 3FD, UK}%
\affiliation{Centre for Theoretical Physics, Department of Physics and Astronomy,
Queen Mary University of London, London E1 4NS, United Kingdom}
\author{Canxin Shi}\email{shicanxin@itp.ac.cn}
\affiliation{Institute of Theoretical Physics,
Chinese Academy of Sciences, Beijing 100190, China}%

\newcommand{\dd}{\mathrm{d}}
\newcommand{\dx}{\mathrm{d} x}
\newcommand{\calO}{\mathcal{O}}
\newcommand{\calE}{\mathcal{E}}

\begin{abstract}
We introduce a new coherent state expansion of the exponential representation of the S-matrix for the classical gravitational two-body problem. By combining the Kosower-Maybee-O'Connell (KMOC) formalism with the Dirac bracket structure emerging in the classical limit, we derive compact and gauge-invariant expressions for scattering observables in the presence of radiation. This causal formulation bypasses the calculation of KMOC cuts and provides a direct link between observables and a minimal set of classical matrix elements extracted from amplitudes. We illustrate our method with several examples, including the impulse, spin kick, angular momentum, waveform and the related radiative fluxes. Finally, using our formalism we evaluate for the first time the spin kick and the change in angular momentum of each particle up to $\mathcal{O}(G^2 s_1^{j_1} s_2^{j_2})$ with $j_1+ j_2  \leq 11$.
\end{abstract}

\maketitle

\section{Introduction}\label{sec:intro}

The detection of gravitational waves from binary sources has opened a new window into the universe, allowing us to probe the nature of black holes and other compact objects with unprecedented precision. The success of this endeavor relies on our ability to provide fast and accurate waveform templates, therefore calling for the development of new theoretical tools to improve the analytic description of the classical two-body problem.

Various perturbative techniques have been developed in the past century to solve Einstein's equations. Among these are the post-Newtonian (PN) expansion for slow-motion, weak-field systems; the post-Minkowskian (PM) expansion for relativistic systems in weak fields; and the self-force (GSF) approach, which expands in the binary’s mass ratio while remaining valid in strong-field regimes. Analytical and numerical data from these methods can be combined together in the powerful effective-one-body (EOB) framework. However, this often involves using gauge-dependent quantities like the Hamiltonian and matching various observables in different regimes.

Focusing on the simplicity on the scattering scenario for generic spinning binaries, we ask here the following questions: what is the \textit{minimal set of gauge-invariant data from which all observables can be derived}? \textit{How can all these observables then be efficiently computed?}

It is now understood that, in the conservative sector, the radial action serves as the generating functional of all binary observables~\cite{Damour:1988mr,Damour:1999cr,Kalin:2019rwq,Kalin:2019inp,Gonzo:2024zxo,Kim:2024svw}. In the scattering case these include the impulse, spin kick, angular momentum (which will be discussed here) as well as the time delay and the elapsed proper time~\cite{Gonzo:2024xjk}. This mirrors the role of the Hamiltonian in classical systems, where the total energy is conserved and the equations of motion determine the corresponding observables. Remarkable progress has been made in recent years on the scattering radial action. At $\mathcal{O}(G)$, it is known to all orders in the spins of both black holes~\cite{Guevara:2018wpp}. At $\mathcal{O}(G^2)$, the current state of the art reaches $\mathcal{O}(G^2 s_1^{j_1} s_2^{j_2})$ with $j_1 + j_2 \leq 11$~\cite{Bohnenblust:2024hkw}. At $\mathcal{O}(G^3)$, for systems with one spinning and one spinless black hole, results are available up to quartic order in spin~\cite{Akpinar:2025bkt}. For a probe spinning particle moving in Kerr spacetime, the expansion is known to all PM orders through $\mathcal{O}(s_1^1 s_2^{\infty})$~\cite{Gonzo:2024zxo}.

In general, though, radiation has to be incorporated in the system and it is unclear what is the natural extension of the radial action. Formally, in the GSF context a gauge-dependent pseudo-Hamiltonian~\cite{Fujita:2016igj,Isoyama:2018sib} -- which include both conservative and dissipative information -- has been defined and recently linked to the first law of black hole mechanics~\cite{Blanco:2022mgd,Gonzo:2024xjk}. Alternatively, an in-in action can be defined~\cite{Galley:2012hx,Jakobsen:2022psy,Kalin:2022hph}, which however requires the doubling of the degrees of freedom and to go beyond the computation of time-ordered matrix elements. To keep working with gauge-invariant amplitudes, one available option is the radiative generalization of the eikonal formalism~\cite{Amati:1987uf,Amati:1990xe,Ciafaloni:2018uwe,Cristofoli:2021jas,DiVecchia:2022piu,DiVecchia:2023frv,Georgoudis:2023lgf,Georgoudis:2023eke}, which is however formulated in a basis which is less convenient for the purpose of this work.

In this paper, we show that the exponential representation of the S-matrix~\cite{Damgaard:2021ipf,Damgaard:2023ttc}, denoted by $\hat{N}$, provides a natural way to define gauge-invariant matrix elements which are classical generating functionals of all scattering observables. In the conservative regime, it has been proven that the four-point matrix element of the $\hat{N}$ operator with external massive particle states is exactly the radial action~\cite{Damgaard:2023ttc,Gonzo:2024xjk}. In section \ref{sec:Noperator} we will develop a coherent state expansion of the $\hat{N}$ operator, where higher-point matrix elements with graviton states are required to define the minimal set of gauge-invariant data. In section \ref{sec:Dirac_brackets} we then combine this novel representation with the KMOC formalism, showing how the recently introduced Dirac brackets and the graviton commutation relations provide a new formula to compute classical scattering observables. Finally, in section \ref{sec:angular_momentum} we apply our findings to compute the spin kick and the angular momentum of spinning bodies at 2PM and to higher orders in spin. In particular, we compute the latter up to the eleventh order in the spin, generalizing the quadratic in spin results~\cite{Jakobsen:2021zvh}.

\paragraph*{Conventions}
For the $(4+N)$-point matrix elements $\mathcal{M}_{4+N}\left(p_1, p_2 ; p_1^{\prime}, p_2^{\prime}, k_1^{\prime}, \ldots, k_N^{\prime}\right)$, we will label the momenta of the incoming (resp. outgoing) massive legs with $p_1^\mu, p_2^\mu$ (resp. $p_1^{\prime \mu}$, $p_2^{\prime \mu}$), while the $N$ outgoing gravitons will have momenta $k_1^{\prime \mu}, \ldots, k_N^{\prime \mu}$. We further define the momentum transfers $q_j^\mu=\hbar \bar{q}_j^\mu$ and the classical 4-velocities $v_j^\mu= \bar{p}_j^\mu/\bar{m}_j$  with $j=1,2$ 
\begin{align}
    &p_1^\mu=\bar{p}_1^\mu+\hbar \frac{\bar{q}_1^\mu}{2}, \quad p_1^{\prime \mu}=\bar{p}_1^\mu-\hbar \frac{\bar{q}_1^\mu}{2}, \nonumber \\
    &p_2^\mu=\bar{p}_2^\mu-\hbar \frac{\bar{q}_2^\mu}{2}, \quad p_2^{\prime \mu}=\bar{p}_2^\mu+\hbar \frac{\bar{q}_2^\mu}{2} \,,
\end{align}
where in the classical limit $\bar{m}_j \to m_j$. We write index symmetrization and anti-symmetrization as $2x^{(\mu} y^{\nu)} = x^\mu y^\nu + x^\nu y^\mu$ and $2x^{[\mu} y^{\nu]} = x^\mu y^\nu - x^\nu y^\mu$, respectively. We also adopt the shorthand notation $\hat{\delta}^n(\cdot)=(2 \pi)^n \delta(\cdot)$ and $\hat{\mathrm{d}}^n q:=\mathrm{d}^n q /(2 \pi)^n$, and we denote the on-shell phase space integral as $\int_k = \int \hat{\mathrm{d}}^3 k/(2 E_k)$. 
Finally, we use the mostly plus $(-+++)$ signature convention for the metric.

\section{Coherent state expansion of the N operator}\label{sec:Noperator}

In this section, we introduce the exponential representation of the S-matrix \cite{Damgaard:2021ipf} in terms of the operator $\hat{N}$
\begin{align}
\label{eq:Noper_def}
\hat{S} = \exp\left(\frac{i}{\hbar} \hat{N} \right)\,,
\end{align}
and we discuss its coherent state expansion for the gravitational two-body scattering problem. Having defined the external scalar and graviton free field quantization
\begin{align}
\label{eq:field_mode}
\phi_j(x) &= \int_p \!\left[\hat{b}_j^{\dagger}(p) e^{-i \frac{p \cdot x}{\hbar}} + \mathrm{h.c.} \right], \qquad j=1,2\,, \nonumber \\
h_{\mu \nu}(x) &= \frac{1}{\sqrt{\hbar}}\sum_{\sigma=\pm}\int_{k} \Big[\varepsilon^{*\sigma}_{\mu \nu}(k) \hat{a}_{\sigma}^{\dagger}(k) e^{-i \frac{k \cdot x}{\hbar}} + \mathrm{h.c.} \Big]\,,
\end{align}
we can expand the $\hat{N}$ operator as \footnote{We added a convenient $\hbar$ normalization for the radiative kernels so that the corresponding tree-level amplitude has the correct classical scaling $\mathcal{M}_{4+N}^{(0)} \stackrel{\hbar \to 0}{\sim} \hbar^{-3-N/2}$, as established in~\cite{Cristofoli:2021jas,Britto:2021pud}.}
\allowdisplaybreaks
\begin{align}
\label{eq:N-exp}
&\hat{N} =  \int_{p_1,p_2,p_1',p_2'} \Big[   (\mathcal{K}) \, \hat{W}^{\mathrm{2-body}}_{p_1',p_2',p_1,p_2} \\
& + \frac{1}{\sqrt{\hbar}} \sum_{\sigma_1}  \int_{k_1'}  (\mathcal{K}_{5, \mathcal{R}}) \, a_{\sigma_1}^{\dagger}(k'_1) \hat{W}^{\mathrm{2-body}}_{p_1',p_2',p_1,p_2} \nonumber \\
& + \frac{1}{2! \hbar} \sum_{\sigma_1,\sigma_2}  \int_{k_1',k_2'} \hspace{-6pt}  (\mathcal{K}^A_{6, \mathcal{R}}) \, a_{\sigma_1}^{\dagger}(k'_1) a_{\sigma_2}^{\dagger}(k'_2) \hat{W}^{\mathrm{2-body}}_{p_1',p_2',p_1,p_2} \nonumber \\
& + \frac{1}{2! \hbar} \sum_{\sigma_1,\sigma_2}  \int_{k_1,k_2'} \hspace{-6pt}  (\mathcal{K}^B_{6, \mathcal{R}}) \, a_{\sigma_2}^{\dagger}(k'_2)  \hat{W}^{\mathrm{2-body}}_{p_1',p_2',p_1,p_2} a_{\sigma_1}(k_1) \!+\! \dots \!+\! \mathrm{h.c.} \Big]\,, \nonumber
\end{align}
in terms of the two-body operator
\begin{align}
\hat{W}^{\mathrm{2-body}}_{p_1',p_2',p_1,p_2} = b_1^{\dagger}(p'_1) b_2^{\dagger}(p'_2) b_1(p_1) b_2(p_2)\,,
\end{align}
and of the following two-massive-particle-irreducible (2MPI) kernels shown in Fig.\ref{fig:2MPI-kernels} \footnote{The definition of $\hat{N}$ matrix elements naturally involve subtracting two-massive-particle reducible contributions from the $\hat{T}$ operator matrix elements, see~\cite{Damgaard:2021ipf,Damgaard:2023ttc} for further details.}
\begin{align}
\label{eq:kernels}
\langle p_1' p_2'| \hat{N} | p_1 p_2 \rangle &= \mathcal{K}\left(p_1,p_2;q\right) \hat{\delta}^4(q_1 + q_2) \,,  \nonumber  \\
\langle p_1' p_2' k_1'| \hat{N} | p_1 p_2 \rangle &=  \mathcal{K}_{5, \mathcal{R}}\left(p_1,p_2; q_1, q_2, k_1'\right) \hat{\delta}^4(q_1 + q_2 - k_1') \,, \nonumber \\
\langle p_1' p_2' k_1' k_2'| \hat{N} | p_1 p_2 \rangle &=  \mathcal{K}^A_{6, \mathcal{R}}\left(p_1,p_2; q_1, q_2, k_1', k_2'\right) \nonumber \\
&\qquad \quad \times \hat{\delta}^4(q_1 + q_2 - k_1'- k_2') \,, \nonumber \\
\langle p_1' p_2' k_2'| \hat{N} | p_1 p_2 k_1\rangle &=  \mathcal{K}^B_{6, \mathcal{R}}\left(p_1,p_2; q_1, q_2, k_1, k_2'\right) \nonumber \\
&\qquad \quad \times \hat{\delta}^4(q_1 + q_2 + k_1 - k_2') \,. 
\end{align}
Note that all the radiative kernels include both connected and disconnected contributions, as we will show later. The kernels $\mathcal{K}^A_{6, \mathcal{R}}$ and $\mathcal{K}^B_{6, \mathcal{R}}$ are evidently related by crossing symmetry, but we label them separately, as establishing this relation can be subtle for higher-point amplitudes~\cite{Caron-Huot:2023ikn}. Moreover, in general $(4+N)$-point kernels $\mathcal{K}_{4+N, \mathcal{R}}$ with $N > 2$ must be introduced, although these are not needed at the order considered in this paper. 

\begin{figure}[t]
\centering

\begin{minipage}{0.47\linewidth}\centering
\begin{tikzpicture}[scale=0.95]
  \draw[kernelline] (-\HalfLen,  \Ysep) -- (\HalfLen,  \Ysep);
  \draw[kernelline] (-\HalfLen, -\Ysep) -- (\HalfLen, -\Ysep);
  \node[anchor=east] at (-\HalfLen-0.08,  \Ysep) {$p_1$};
  \node[anchor=east] at (-\HalfLen-0.08, -\Ysep) {$p_2$};
  \node[anchor=west, yshift=0.5mm] at ( \HalfLen+0.12,  \Ysep) {$p'_1$};
  \node[anchor=west, yshift=0.5mm] at ( \HalfLen+0.12, -\Ysep) {$p'_2$};
  \node[blobstyle] (A) at (0,0) {};                
  \node at ([yshift=0.35mm]A.center) {\Large $\hat N$};
  \node[below=4pt] at (A.south) {$\mathcal K_4$};
\end{tikzpicture}
\end{minipage}\hfill
\begin{minipage}{0.47\linewidth}\centering
\begin{tikzpicture}[scale=0.95]
  \draw[kernelline] (-\HalfLen,  \Ysep) -- (\HalfLen,  \Ysep);
  \draw[kernelline] (-\HalfLen, -\Ysep) -- (\HalfLen, -\Ysep);
  \node[anchor=east] at (-\HalfLen-0.08,  \Ysep) {$p_1$};
  \node[anchor=east] at (-\HalfLen-0.08, -\Ysep) {$p_2$};
  \node[anchor=west, yshift=0.5mm] at ( \HalfLen+0.12,  \Ysep) {$p'_1$};
  \node[anchor=west, yshift=0.5mm] at ( \HalfLen+0.12, -\Ysep) {$p'_2$};
  \draw[gluon] (0.55,0) -- (1.55,0);
  \node[anchor=west, yshift=0.25mm] at (1.60,0) {$k'_1$};
  \node[below=4pt] at (A.south) {$\mathcal K_{5,\mathcal R}$};
  \node[blobstyle] (A) at (0,0) {};                
  \node at ([yshift=0.35mm]A.center) {\Large $\hat N$};
\end{tikzpicture}
\end{minipage}

\vspace{0.9em}

\begin{minipage}{0.47\linewidth}\centering
\begin{tikzpicture}[scale=0.95]
  \draw[kernelline] (-\HalfLen,  \Ysep) -- (\HalfLen,  \Ysep);
  \draw[kernelline] (-\HalfLen, -\Ysep) -- (\HalfLen, -\Ysep);
  \node[anchor=east] at (-\HalfLen-0.08,  \Ysep) {$p_1$};
  \node[anchor=east] at (-\HalfLen-0.08, -\Ysep) {$p_2$};
  \node[anchor=west, yshift=0.5mm] at ( \HalfLen+0.12,  \Ysep) {$p'_1$};
  \node[anchor=west, yshift=0.5mm] at ( \HalfLen+0.12, -\Ysep) {$p'_2$};
  \draw[gluon] (1.55,  0.20) -- (0.55,  0.20);
  \draw[gluon] (1.55, -0.20) -- (0.55, -0.20);
  \node[anchor=west, yshift=0.25mm] at (1.60,  0.20) {$k'_1$};
  \node[anchor=west, yshift=0.25mm] at (1.60, -0.20) {$k'_2$};
  \node[below=4pt] at (A.south) {$\mathcal K^{A}_{6,\mathcal R}$};
  \node[blobstyle] (A) at (0,0) {};                
  \node at ([yshift=0.35mm]A.center) {\Large $\hat N$};
\end{tikzpicture}
\end{minipage}\hfill
\begin{minipage}{0.47\linewidth}\centering
\begin{tikzpicture}[scale=0.95]
  \draw[kernelline] (-\HalfLen,  \Ysep) -- (\HalfLen,  \Ysep);
  \draw[kernelline] (-\HalfLen, -\Ysep) -- (\HalfLen, -\Ysep);
  \node[anchor=east] at (-\HalfLen-0.08,  \Ysep) {$p_1$};
  \node[anchor=east] at (-\HalfLen-0.08, -\Ysep) {$p_2$};
  \node[anchor=west, yshift=0.5mm] at ( \HalfLen+0.08,  \Ysep) {$p'_1$};
  \node[anchor=west, yshift=0.5mm] at ( \HalfLen+0.08, -\Ysep) {$p'_2$};
  \draw[gluon] (-1.55, 0.00) -- (-0.55, 0.00);
  \draw[gluon] ( 0.55, 0.00) -- ( 1.55, 0.00);
  \node[anchor=east, yshift=0.25mm] at (-1.60, 0.00) {$k_1$};
  \node[anchor=west, yshift=0.25mm] at ( 1.60, 0.00) {$k'_2$};
  \node[below=4pt] at (A.south) {$\mathcal K^{B}_{6,\mathcal R}$};
  \node[blobstyle] (A) at (0,0) {};                
  \node at ([yshift=0.35mm]A.center) {\Large $\hat N$};
\end{tikzpicture}
\end{minipage}

\caption{Relevant 2MPI kernels considered in this work.}
\label{fig:2MPI-kernels}
\end{figure}

We now study the evolution of the state $\mid \!\!\text{in} \rangle = \mid \!\!p_1 p_2 0 \rangle$ of two massive scalar particles initially separated by the impact parameter $b^\mu = b_2^{\mu} - b_1^{\mu}$. We can simplify the action of the $\hat{S}$-matrix \eqref{eq:Noper_def} on such state by working in impact parameter space, where the superclassical iterations manifestly exponentiate in the heavy-mass expansion~\cite{Cristofoli:2021jas,Bjerrum-Bohr:2021wwt,DiVecchia:2022piu,Brandhuber:2021eyq,Brandhuber:2023hhy,Adamo:2024oxy}. 
Introducing the Fourier conjugate representation of the kernels in the classical regime
\begin{align}
\label{eq:Fourier_transform}
&\!\!\widetilde{\mathcal{K}}^{\mathrm{cl}}\left(b\right) = \int  \hat{\mathrm{d}}^4 q \,\hat{\delta}\left(2 \bar{p}_1 \cdot q\right) \hat{\delta}\left(2 \bar{p}_2 \cdot q\right) \mathrm{e}^{i\left(q \cdot b\right) / \hbar} \mathcal{K}^{\mathrm{cl}}\left(q\right)\,, \\
&\!\!\widetilde{\mathcal{K}}_{4+N, \mathcal{R}}^{\mathrm{cl}}\left(b_1, b_2; \{k_j\}_{j=1,\dots N}\right) \nonumber \\
&\!= \int  \hat{\mathrm{d}}^4 q_1 \hat{\mathrm{d}}^4 q_2 \,  \hat{\delta}\left(2 \bar{p}_1 \cdot q_1\right) \hat{\delta}\left(2 \bar{p}_2 \cdot q_2\right)  \hat{\delta}^4\Big(q_1+q_2-\sum_{j=1}^N k_j\Big) \nonumber \\
& \qquad \times\mathrm{e}^{i\left(q_1 \cdot b_1+q_2 \cdot b_2\right) / \hbar} \mathcal{K}_{4+N, \mathcal{R}}^{\mathrm{cl}}\left(q_1, q_2; \{k_j\}_{j=1,\dots N}\right)\,,  \nonumber 
\end{align}
we can equivalently write the action on the state as \footnote{While this has not been proven to all orders in perturbation theory, explicit calculations both for the 4-point~\cite{Brandhuber:2023hhy} and 5-point~\cite{Brandhuber:2023hhy} amplitude at loop order fully support this conjecture.}
\begin{align}
\label{eq:S-matrix_eff}
&\exp\left(\frac{i}{\hbar} \hat{N} \right) |\text{in}\rangle \stackrel{\hbar \to 0}{\sim} \hat{S}^{\mathrm{cl}}  |\bar{p}_1 \bar{p}_2\rangle\,,
\end{align}
as a coherent state expansion 
\begin{align}
\label{eq:HEFT_exp}
& \hat{S}^{\mathrm{cl}}=\exp\Big\{\frac{i}{\hbar} \Big[\widetilde{\mathcal{K}}^{\mathrm{cl}}\left(b\right) + \sum_{\sigma_1} \frac{1}{\sqrt{\hbar}} \int_{k_1^{\prime}} \hat{\alpha}_{5, \mathcal{R}}^{\mathrm{cl}}\left(k_1^{\prime}\right)  \\
& \quad+ \sum_{N=2}^{+\infty} \frac{1}{N! \hbar^{N/2}} \sum_{\sigma_1, \ldots, \sigma_N}  \int_{k_1^{\prime},\dots k_N^{\prime}} \hat{\alpha}_{4+N, \mathcal{R}}^{\mathrm{cl}}\left(k_1^{\prime}, \ldots, k_N^{\prime}\right)\Big] \Big\}\,,  \nonumber
\end{align}
in terms of the operators $\hat{\alpha}_{4+N, \mathcal{R}}^{\mathrm{cl}}$ which encode the contributions of $N$ graviton emissions. The first radiative operator with one graviton mode (i.e, $N=1$) corresponds to a single coherent state~\cite{Cristofoli:2021jas,Britto:2021pud,DiVecchia:2022piu}
\begin{align}
\label{eq:alphaN1}
& i\, \hat{\alpha}_{5, \mathcal{R}}^{\mathrm{cl}}\left(k_1^{\prime}\right)=i\, \widetilde{\mathcal{K}}_{5, \mathcal{R}}^{\mathrm{cl}}\left(b_1, b_2;k_1^{\prime}\right) a_{\sigma_1}^{\dagger}\left(k_1^{\prime}\right)  \\
& \qquad \qquad \qquad \qquad -(i \,\widetilde{\mathcal{K}}_{5, \mathcal{R}}^{ \mathrm{cl}}\left(b_1, b_2; k_1^{\prime}\right))^* a_{\sigma_1}\left(k_1^{\prime}\right)\,,\nonumber 
\end{align}
while the higher order operators $N > 1$ correspond to deviations from a Poissonian distribution. Up to the perturbative order considered in this work, only the $N=2$ operator -- describing with \eqref{eq:alphaN1} a two-mode coherent squeezed state~\cite{Lee1990,Caves1991} -- is potentially relevant \footnote{See also~\cite{Fernandes:2024xqr,Aoude:2024sve} for other interesting works in this direction.}
\begin{align}
\label{eq:alphaN2}
& \hspace{-10pt}\hat{\alpha}_{6, \mathcal{R}}^{\mathrm{cl}}\left(k_1^{\prime},k_2^{\prime}\right) =  \widetilde{\mathcal{K}}_{6, \mathcal{R}}^{A\mathrm{cl}}\left(b_1, b_2; k_1^{\prime},k_2^{\prime}\right) a_{\sigma_1}^{\dagger}\left(k_1^{\prime}\right) a_{\sigma_2}^{\dagger}\left(k_2^{\prime}\right)  \\
& \quad \qquad \qquad \quad +\widetilde{\mathcal{K}}_{6, \mathcal{R}}^{A* \mathrm{cl}}\left(b_1, b_2; k_1^{\prime}, k_2^{\prime}\right) a_{\sigma_1}\left(k_1^{\prime}\right) a_{\sigma_2}\left(k_2^{\prime}\right) \nonumber \\
& \quad \qquad \qquad \quad +\widetilde{\mathcal{K}}_{6, \mathcal{R}}^{B \mathrm{cl}}\left(b_1, b_2; k_1, k_2^{\prime}\right)  a^{\dagger}_{\sigma_2}\left(k_2^{\prime}\right) a_{\sigma_1}\left(k_1\right) \nonumber \\
& \quad \qquad \qquad \quad +\widetilde{\mathcal{K}}_{6, \mathcal{R}}^{B* \mathrm{cl}}\left(b_1, b_2; k_1, k_2^{\prime}\right)  a^{\dagger}_{\sigma_1}\left(k_1\right) a_{\sigma_2}\left(k_2^{\prime}\right) \,,\nonumber
\end{align}
although it has been shown to be quantum suppressed in the analysis of the variance~\cite{Cristofoli:2021jas,Britto:2021pud} and the impulse~\cite{Damgaard:2023ttc} at the first order in which it appears, i.e. at 4PM. Here we adopt a cautious approach and retain it explicitly, although the classical scaling imposed in \eqref{eq:HEFT_exp} will force some contributions to vanish. The generalization of \eqref{eq:alphaN1} and \eqref{eq:alphaN2} to the case of $N$-graviton operator $\hat{\alpha}_{4+N, \mathcal{R}}^{\mathrm{cl}}\left(k_1^{\prime},k_2^{\prime}\right)$  is straightforward, and will not be discussed further here. 

Finally, it is easy to see that the proposed coherent state expansion of the S-matrix in \eqref{eq:HEFT_exp}
\begin{align}
\label{eq:HEFT_new}
& \hat{S}^{\mathrm{cl}}=\exp\Big\{\frac{i}{\hbar} \Big[ \widetilde{\mathcal{K}}^{\mathrm{cl}} + \sum_{N=1}^{+\infty} \hat{\alpha}_{4+N, \mathcal{R}}^{\mathrm{cl}} \Big]\Big\}\,,  \\
&\hat{\alpha}_{4+N, \mathcal{R}}^{\mathrm{cl}} := \frac{1}{N! \hbar^{N/2}} \sum_{\sigma_1,\dots,\sigma_N} \int_{k_1^{\prime},\dots, k_N^{\prime}} \hat{\alpha}_{4+N, \mathcal{R}}^{\mathrm{cl}}\left(k_1^{\prime},\dots, k_N^{\prime}\right)\,, \nonumber 
\end{align}
provides a unitary representation of the dynamics.

\section{Dirac brackets for all scattering observables}\label{sec:Dirac_brackets}

In the novel representation \eqref{eq:HEFT_new}, which is valid also for the spinning case, the kinematic phase space of the two-body problem is 
\begin{align}
    \label{eq:phase_space}
    & \qquad \qquad \qquad \qquad \mathcal{H} = \mathcal{H}_{\mathrm{2-body}} \cup \mathcal{H}_{\mathcal{R}}\,, \\
    &\mathcal{H}_{2-\mathrm{body}} = \{v_1,v_2,s_1,s_2,b_1,b_2\}\,, \,\,\, \mathcal{H}_{\mathcal{R}}= \{k_1,\dots k_N\} \,.  \nonumber 
\end{align}
We consider the variation of a classical observable $\hat{O}$ \cite{Kosower:2018adc},
\begin{align}
    \Delta \hat{O}=(\hat{S}^{\mathrm{cl}})^{\dagger} \hat{O} \hat{S}^{\mathrm{cl}}-\hat{O}\,,
    \label{eq:observables}
\end{align}
and we use the Campbell identity 
\begin{align}
    e^{\hat{U}} \hat{O} e^{-\hat{U}} = \sum_{n = 0}^{+\infty} \frac{1}{n!} \underbrace{[\hat{U} ,[\hat{U} , \ldots,[\hat{U}}_{\mathrm{n} \text { times }} , \hat{O}]]]\,,
\end{align}
which can be conveniently recasted in terms of the iterated commutator
\begin{align}
\label{eq:nested_commut1}
   & \qquad \qquad e^{\hat{U}} \hat{O} e^{-\hat{U}} = \sum_{n = 0}^{+\infty} \frac{1}{n!} [(\hat{U} )^{\odot n}, \hat{O}]\,,  \\
   & [(\hat{U} )^{\odot n}, \hat{O}] := \underbrace{[\hat{U} ,[\hat{U} , \ldots,[\hat{U} }_{\mathrm{n} \text { times }}, \hat{O}]]]\,, \quad  [(\hat{U} )^{\odot 0}, \hat{O}] = \hat{O}\,. \nonumber
\end{align}
Applying \eqref{eq:nested_commut1} iteratively to \eqref{eq:observables}, we obtain
\begin{align}
\label{eq:master_formula}
   & \Delta \hat{O} =  \sum_{n = 1}^{+\infty} \frac{(-i)^{n}}{\hbar^{n}} \frac{1}{n!} \Big[\Big(\widetilde{\mathcal{K}}^{\mathrm{cl}} + \sum_{j=1}^N \hat{\alpha}_{4+j, \mathcal{R}}\Big)^{\odot n}, \hat{O}\Big]\,,
\end{align}
which is an important result for classical observables. Note, in particular, that \eqref{eq:nested_commut1} allows to write the contributions of the form $T^{\dagger} \hat{O} T$ coming from the expansion $S = 1 + i T$  -- also called KMOC cut-- in a completely new fashion, as also emphasized in \cite{Damgaard:2023ttc,Kim:2025hpn}. We will now show that the classical limit of the nested commutators in \eqref{eq:master_formula} gives an iterative master formula for the calculation of a generic observables from the classical phase space.

In general, the operator definition of radiative ($\hat{O}_{\mathrm{R}}$) and two-body ($\hat{O}_{\mathrm{2-body}}$) observables involves either classical kinematic variables or the graviton mode oscillators
\begin{align}
    \hat{O}_{\mathrm{2-body}}& = f_{\mathcal{O}}(v_1,v_2,s_1,s_2,b_1,b_2)\,, \nonumber \\
    &\quad \hat{O}_{\mathrm{R}} = g_{\mathcal{O}}(a,a^{\dagger})\,,
\end{align}
where $f_{\mathcal{O}}$ and $g_{\mathcal{O}}$ have a polynomial dependence on their arguments, but are otherwise generic. When computing the commutators, we use the free graviton algebra\footnote{While it is possible to promote this to an algebra of the graviton field \cite{Kim:2025hpn}, we find it more convenient to have a mixed representation in the spirit of the eikonal formulation.}
\begin{align}
    [a(k), a^{\dagger}(k')] = 2 E_k \, \hat{\delta}^3(k - k')\,,
    \label{eq:free_graviton}
\end{align}
and also the classical Dirac brackets \cite{Gonzo:2024zxo}
\begin{align}
    [f (\cdot), g (\cdot) ] \to i \hbar  \{ f (\cdot), g (\cdot)\}_{\DB}\,.
    \label{eq:Dirac_brackets}
\end{align}
Notice that in general we will have contributions both from \eqref{eq:Dirac_brackets} and from \eqref{eq:free_graviton}. The structure of the Dirac brackets follows from the quantization of a relativistic system of interacting spinning point particles, after imposing a set of constraints on the usual Poisson brackets 
\begin{gather}
\label{eq:Poisson_brackets}
    \{b_i^\mu, v_j^\nu\}_{\PB} = \delta_{ij} \frac{\eta^{\mu\nu}}{m_i},
    \\
    \{S_i^{\mu \nu}, S_j^{\alpha \beta}\}_{\PB}= \delta_{ij} \left[ S_i^{\mu \alpha}\eta^{\nu \beta} - S_i^{\mu \beta} \eta^{\nu \alpha} - (\mu \leftrightarrow \nu) \right]\,,
    \nonumber
\end{gather}
valid for both particles $i = 1,2$ with their spin tensor $S_i^{\mu\nu}$. Then, if the spin supplementary condition (SSC) $v_{i, \mu} S_i^{\mu\nu} = 0$ is imposed, the Poisson brackets \eqref{eq:Poisson_brackets} become the primed brackets \cite{Hanson:1974qy}
\begin{align}
\label{eq:primed_brackets}
    & \{b_i^\mu, v_j^\nu\}^{\prime} = \delta_{ij} \frac{\eta^{\mu\nu}}{m_j}\,, \qquad\, \{b_i^\mu, b_j^\nu\}^{\prime} = \delta_{ij} \frac{S_i^{\mu\nu}}{m_i^2}\,, \\
    &\{b_i^\mu, s_j^\nu\}^{\prime} = \delta_{ij} \frac{2 s_i^{[\mu}v_i^{\nu]}}{m_j}\,, \qquad \{s_i^\mu, s_j^\nu\}^{\prime} = \delta_{ij}  \frac{S_i^{\mu\nu}}{m_i^2}\,, \nonumber
\end{align}
where we have defined $S_i^{\mu\nu} = m_i \epsilon^{\mu\nu}{}_{\rho\sigma} v_i^\rho s_i^{\sigma}$.

Finally, after imposing also the on-shellness  $v_{j} \cdot v_{j} = -1$ and the transversality constraints $(b_2 - b_1) \cdot v_{j} = 0$ for $j=1,2$ we obtain the Dirac brackets \cite{Gonzo:2024zxo}
\begin{subequations}
\label{eq:Dirac_brackets_final}
\begin{gather}
    \{b_i^\mu, v_j^\nu\}_{\DB} = \delta_{ij} \frac{\eta^{\mu\nu}}{m_j} + (-1)^{\delta_{ij}} \frac{\mathcal{P}_i^{\mu\nu}}{m_j}, \\
    \{b_i^\mu, s_j^\nu\}_{\DB} = \delta_{ij} \frac{s_i^\mu v_i^\nu}{m_j} + (-1)^{\delta_{ij}} \frac{\mathcal{P}_i^{\mu\rho}s_{j\rho} v_j^\nu}{m_j}, \\
    \{b_i^\mu, b_j^\nu\}_{\DB} = \frac{v_i \cdot v_j}{\sigma^2 -1} \left( \mathrm{sgn}_i \frac{b^\mu v_j^\nu}{m_i} -  \mathrm{sgn}_j \frac{b^\nu v_i^\mu}{m_j}\right) \\
    \quad+ \delta_{ij} \frac{S_i^{\mu\nu}}{m_i^2} + (-1)^{\delta_{ij}} \left(\mathcal{P}_i^{\mu\rho} \frac{S_{j\rho}{}^{\nu}}{m_j^2} - \mathcal{P}_j^{\nu\rho} \frac{S_{i\rho}{}^{\mu}}{m_i^2}\right) ,\nonumber\\
    \{s_i^\mu, s_j^\nu\}_{\DB} = \delta_{ij}  \frac{S_i^{\mu\nu}}{m_i^2}, 
\end{gather}
\end{subequations}
where we have defined $\sigma := -v_1 {\cdot} v_2$, and 
\begin{gather}
    \mathrm{sgn}_1 := -1, \quad 
    \mathrm{sgn}_2 := +1, \\ 
    \mathcal{P}_1^{\mu\nu} := \frac{v_1^\mu (v_1^\nu - \sigma v_2^\nu)}{\sigma^2 -1},\quad 
    \mathcal{P}_2^{\mu\nu} := \frac{v_2^\mu (v_2^\nu - \sigma v_1^\nu)}{\sigma^2 -1}.\quad 
\end{gather}

\vspace{-14pt}

\subsection{Impulse, spin kick and angular momentum}

\vspace{-6pt}

We now apply our formalism to the calculation of two-body observables $\hat{O}_{\mathrm{2-body}}$, such as the impulse, the impact parameter shift and the spin kick for generic spinning binaries. Denoting $\lambda^{\mu} = \{v_1^{\mu},v_2^{\mu},s_1^{\mu},s_2^{\mu},b_1^{\mu},b_2^{\mu}\}$, we obtain the following expression valid up to 5PM order 
\allowdisplaybreaks
\begin{align}
    \label{eq:impulse}
    &\langle \Delta \lambda^{\mu}  \rangle \Big|_{\mathcal{O}(G^5)} =  \sum_{n=1}^{+\infty} \frac{1}{n!} \big\{ (\widetilde{\mathcal{K}}^{\mathrm{cl}})^{\odot n}, \lambda^\mu \big\}_{\DB} \nonumber \\
    & - \frac{i}{2!} \int_k \Big(\widetilde{\mathcal{K}}_{5, \mathcal{R}}^{*\mathrm{cl}}(k) \{ \widetilde{\mathcal{K}}_{5, \mathcal{R}}^{\mathrm{cl}}(k), \lambda^{\mu}\}_{\DB} -\mathrm{c.c} \Big) \nonumber\\
    & - \frac{i}{3!} \int_k \Big(\widetilde{\mathcal{K}}_{5, \mathcal{R}}^{*\mathrm{cl}}(k) \{ \widetilde{\mathcal{K}}_{5, \mathcal{R}}^{\mathrm{cl}}(k),\{ \widetilde{\mathcal{K}}^{\mathrm{cl}}, \lambda^{\mu}\}_{\DB}\}_{\DB} - \mathrm{c.c.}  \nonumber \\
    & \qquad \quad +\widetilde{\mathcal{K}}_{5, \mathcal{R}}^{*\mathrm{cl}}(k) \{ \widetilde{\mathcal{K}}^{\mathrm{cl}},\{\widetilde{\mathcal{K}}_{5, \mathcal{R}}^{\mathrm{cl}}(k) , \lambda^{\mu}\}_{\DB}\}_{\DB} - \mathrm{c.c.}  \nonumber \\
    & \qquad \quad + \{ \widetilde{\mathcal{K}}^{\mathrm{cl}}, \widetilde{\mathcal{K}}_{5, \mathcal{R}}^{*\mathrm{cl}}(k)\{\widetilde{\mathcal{K}}_{5, \mathcal{R}}^{\mathrm{cl}}(k) , \lambda^{\mu}\}_{\DB}\}_{\DB} - \mathrm{c.c.} \Big) \nonumber \\
    &- \frac{i}{4!} \int_k \Big( \widetilde{\mathcal{K}}_{5, \mathcal{R}}^{*\mathrm{cl}}(k) \{ \widetilde{\mathcal{K}}_{5, \mathcal{R}}^{\mathrm{cl}}(k), \{ (\widetilde{\mathcal{K}}^{\mathrm{cl}})^{\odot2}, \lambda^\mu \}_{\DB}  \}_{\DB} - \mathrm{c.c.}
    \nonumber  \\
    & \qquad\quad + \widetilde{\mathcal{K}}_{5, \mathcal{R}}^{*\mathrm{cl}}(k) \{ (\widetilde{\mathcal{K}}^{\mathrm{cl}})^{\odot2},\{ \widetilde{\mathcal{K}}_{5, \mathcal{R}}^{\mathrm{cl}}(k), \lambda^\mu \}_{\DB}  \}_{\DB} - \mathrm{c.c.} 
    \nonumber  \\
    & \qquad\quad +  \{ (\widetilde{\mathcal{K}}^{\mathrm{cl}})^{\odot2}, \widetilde{\mathcal{K}}_{5, \mathcal{R}}^{*\mathrm{cl}}(k)\{ \widetilde{\mathcal{K}}_{5, \mathcal{R}}^{\mathrm{cl}}(k), \lambda^\mu \}_{\DB}  \}_{\DB} - \mathrm{c.c.} 
    \nonumber  \\
    & \qquad +\widetilde{\mathcal{K}}_{5, \mathcal{R}}^{*\mathrm{cl}}(k)  \{ \widetilde{\mathcal{K}}^{\mathrm{cl}}, \{ \widetilde{\mathcal{K}}_{5, \mathcal{R}}^{\mathrm{cl}}(k), \{ \widetilde{\mathcal{K}}^{\mathrm{cl}}, \lambda^\mu \}_{\DB} \}_{\DB} \}_{\DB} - \mathrm{c.c.} 
    \nonumber  \\
    & \qquad + \{ \widetilde{\mathcal{K}}^{\mathrm{cl}}, \widetilde{\mathcal{K}}_{5, \mathcal{R}}^{*\mathrm{cl}}(k) \{ \widetilde{\mathcal{K}}_{5, \mathcal{R}}^{\mathrm{cl}}(k), \{ \widetilde{\mathcal{K}}^{\mathrm{cl}}, \lambda^\mu \}_{\DB} \}_{\DB} \}_{\DB} - \mathrm{c.c.} 
    \nonumber  \\
    & \qquad + \{ \widetilde{\mathcal{K}}^{\mathrm{cl}}, \widetilde{\mathcal{K}}_{5, \mathcal{R}}^{*\mathrm{cl}}(k) \{ \widetilde{\mathcal{K}}^{\mathrm{cl}}, \{ \widetilde{\mathcal{K}}_{5, \mathcal{R}}^{\mathrm{cl}}(k), \lambda^\mu \}_{\DB} \}_{\DB} \}_{\DB} - \mathrm{c.c.} \Big)
    \nonumber  \\
    & + \frac{1}{3!} \int_{k_1,k_2} \hspace{-6pt} \Big(  \{ \widetilde{\mathcal{K}}_{5, \mathcal{R}}^{\mathrm{cl}}(k_1), \lambda^\mu \}_{\DB} \, \widetilde{\mathcal{K}}_{5, \mathcal{R}}^{\mathrm{cl}}(k_2) \, \widetilde{\mathcal{K}}_{6, \mathcal{R}}^{A*\mathrm{cl}}(k_1,k_2)  \nonumber \\
    & \qquad \qquad + \{ \widetilde{\mathcal{K}}_{5, \mathcal{R}}^{*\mathrm{cl}}(k_1), \lambda^\mu \}_{\DB} \, \widetilde{\mathcal{K}}_{5, \mathcal{R}}^{*\mathrm{cl}}(k_2) \, \widetilde{\mathcal{K}}_{6, \mathcal{R}}^{A\mathrm{cl}}(k_1,k_2) \nonumber \\
    &\qquad\qquad - \{ \widetilde{\mathcal{K}}_{6, \mathcal{R}}^{A*\mathrm{cl}}(k_1,k_2), \lambda^\mu \}_{\DB} \, \widetilde{\mathcal{K}}_{5, \mathcal{R}}^{\mathrm{cl}}(k_1) \, \widetilde{\mathcal{K}}_{5, \mathcal{R}}^{\mathrm{cl}}(k_2)  \nonumber \\
    & \qquad \qquad - \{ \widetilde{\mathcal{K}}_{6, \mathcal{R}}^{A\mathrm{cl}}(k_1,k_2), \lambda^\mu \}_{\DB} \, \widetilde{\mathcal{K}}_{5, \mathcal{R}}^{*\mathrm{cl}}(k_1) \, \widetilde{\mathcal{K}}_{5, \mathcal{R}}^{*\mathrm{cl}}(k_2) \Big)\nonumber \\
    &  + \frac{1}{3!}  \int_{k_1,k_2} \hspace{-6pt} \Big( 
    \{ \widetilde{\mathcal{K}}_{6, \mathcal{R}}^{B*\mathrm{cl}}(k_1,k_2), \lambda^\mu \}_{\DB} \, \widetilde{\mathcal{K}}_{5, \mathcal{R}}^{\mathrm{cl}}(k_1) \, \widetilde{\mathcal{K}}_{5, \mathcal{R}}^{*\mathrm{cl}}(k_2) \nonumber \\
    & \qquad \qquad + \{ \widetilde{\mathcal{K}}_{6, \mathcal{R}}^{B\mathrm{cl}}(k_1,k_2), \lambda^\mu \}_{\DB} \, \widetilde{\mathcal{K}}_{5, \mathcal{R}}^{\mathrm{cl}}(k_1) \, \widetilde{\mathcal{K}}_{5, \mathcal{R}}^{*\mathrm{cl}}(k_2) 
    \Big) \nonumber \\
    & - \frac{2}{4!}  \int_{k_1,k_2} \hspace{-6pt}\Big(
    \widetilde{\mathcal{K}}_{6, \mathcal{R}}^{B\mathrm{cl}}(k_1,k_2) + \widetilde{\mathcal{K}}_{6, \mathcal{R}}^{B*\mathrm{cl}}(k_1,k_2) 
    \Big) \nonumber \\
    &  \qquad  \qquad \times \Big( 
    \{ \widetilde{\mathcal{K}}_{5, \mathcal{R}}^{*\mathrm{cl}}(k_1), \lambda^\mu \}_{\DB} \, \widetilde{\mathcal{K}}_{5, \mathcal{R}}^{\mathrm{cl}}(k_2) 
    \nonumber \\
    & \qquad \qquad  \qquad  \qquad + \{ \widetilde{\mathcal{K}}_{5, \mathcal{R}}^{\mathrm{cl}}(k_1), \lambda^\mu \}_{\DB} \, \widetilde{\mathcal{K}}_{5, \mathcal{R}}^{*\mathrm{cl}}(k_2) 
    \Big) \,,
\end{align}
which is a novel compact representation of generic observables in terms of gauge-invariant matrix elements that can be extracted from amplitude calculations. For compactness, we omitted the sum over helicity states of the intermediate gravitons. 

A striking resummed pattern emerges in \eqref{eq:impulse}. Including only $\widetilde{\mathcal{K}}^{\mathrm{cl}}$ and a single pair $(\widetilde{\mathcal{K}}_{5, \mathcal{R}}^{*\mathrm{cl}},\widetilde{\mathcal{K}}_{5, \mathcal{R}}^{\mathrm{cl}})$, and neglecting the higher-order radiative kernels $\widetilde{\mathcal{K}}_{4+N, \mathcal{R}}^{\mathrm{cl}}$ with $N\geq 2$, we obtain the simpler closed-form expression
\begin{align}
    \label{eq:impulse_res}
    &\langle \Delta \lambda^{\mu}  \rangle \sim  \sum_{n=1}^{+\infty} \frac{1}{n!} \big\{ (\widetilde{\mathcal{K}}^{\mathrm{cl}})^{\odot n}, \lambda^\mu \big\}_{\DB}  \\
    &-\sum_{n=2}^{+\infty} \frac{i}{n!}\!\! \sum_{\substack{a,b,c\geq0\\a+b+c=n{-}2}} \int_k  \Big( \big\{ (\widetilde{\mathcal{K}}^{\mathrm{cl}})^{\odot a},\widetilde{\mathcal{K}}_{5, \mathcal{R}}^{*\mathrm{cl}}(k)  \! \nonumber\\
    &\!\times  
     \big\{ (\widetilde{\mathcal{K}}^{\mathrm{cl}})^{\odot b},  \big\{\widetilde{\mathcal{K}}_{5, \mathcal{R}}^{\mathrm{cl}}(k) ,
    \big\{(\widetilde{\mathcal{K}}^{\mathrm{cl}})^{\odot c}, \lambda^\mu
    \big\}_{\DB} \big\}_{\DB} \big\}_{\DB} \big\}_{\DB}- \mathrm{c.c.} \Big)\,, \nonumber
\end{align}
which suggests the possibility of resumming \eqref{eq:impulse} into a more compact expression. We also note that several terms vanish as they are purely quantum, such as
\begin{align}
&\hbar \int_k \,\{\widetilde{\mathcal{K}}_{5, \mathcal{R}}^{\mathrm{cl}}(k), \{ \widetilde{\mathcal{K}}_{5, \mathcal{R}}^{*\mathrm{cl}}, \lambda^{\mu}\}_{\DB} \}_{\DB} \propto \mathcal{O}(\hbar)\,, \nonumber \\
&\hbar \int_{k_1,k_2} \widetilde{\mathcal{K}}_{6, \mathcal{R}}^{A\mathrm{cl}} \{ \widetilde{\mathcal{K}}_{6, \mathcal{R}}^{A*\mathrm{cl}}, \lambda^{\mu}\}_{\DB} \propto \mathcal{O}(\hbar)\,.
\end{align}
The latter suppression, established in \cite{Cristofoli:2021jas,Britto:2021pud, Damgaard:2023ttc}, is a direct consequence of the fact that pure graviton loops give quantum effects.

Using the linearity of the Dirac brackets, we can also compute the variation of the angular momentum for each particle with the formula \eqref{eq:impulse}. Indeed, the angular momentum consists of a mechanical and spin contribution 
\begin{align}
\label{eq:angular_momentum}
    J_i^{\mu \nu}&=2 m_i b_i^{[\mu} v_i^{\nu]}+S_i^{\mu\nu}\,, \quad i=1,2\,.
\end{align}
whose variation depends only on the changes in position $\langle \Delta b_i^{\mu} \rangle$, velocity $\langle \Delta v_i^{\nu} \rangle$ and spin $\langle \Delta s_i^{\sigma} \rangle$:
\begin{align}
\label{eq:Ji_expansion}
   \langle \Delta J_i^{\mu\nu} \rangle &=2 m_i (b_i^{[\mu} + \langle \Delta b_i^{[\mu} \rangle) (v_i^{\nu]} + \langle \Delta v_i^{\nu]} \rangle)  \\
   &+ m_i \epsilon^{\mu\nu}{}_{\rho\sigma} (v_i^\rho +  \langle \Delta v_i^\rho \rangle) (s_i^{\sigma} + \langle \Delta s_i^{\sigma} \rangle) \quad i=1,2\,.\nonumber
\end{align}

It is worth making few remarks about \eqref{eq:impulse}. First, in the conservative case $\widetilde{\mathcal{K}}_{4+j, \mathcal{R}}^{\mathrm{cl}} \to 0, \, \forall j \geq 1$, and the special kinematics allow to uniquely identify $\widetilde{\mathcal{K}}^{\mathrm{cl}}\left(b\right)$ with the radial action $I_r(b)$ \cite{Damgaard:2023ttc}, confirming earlier expectations \cite{Bern:2021dqo,Brandhuber:2021eyq,Bjerrum-Bohr:2021wwt,Kol:2021jjc,Adamo:2022ooq}. Therefore, in such case only the first line of \eqref{eq:impulse} survives and it reduces to eq.(37) of \cite{Gonzo:2024zxo}. We also notice that the 1PM spinless piece of the radial action is infrared divergent. For $\langle\Delta v_i^{\mu}\rangle$ and $\langle\Delta s_i^{\mu}\rangle$ this is not problematic, since the Dirac bracket with the regulator-dependent part simply vanishes. However, for $\Delta b_i^\mu$, and consequently for $\Delta J_i^{\mu\nu}$, we obtain a non-trivial divergent result. To deal with this problem, we subtract the leading order divergence from the radial action, defining
\begin{align}
    \widetilde{\mathcal{K}}^{\mathrm{cl}}\big|_{\mathcal{O}(G s_1^0s_2^0)} \equiv -\frac{2 G m_1 m_2 \left(2 \sigma ^2-1\right) \log |b|}{\sqrt{\sigma ^2-1}}\,.
    \label{eq:subtraction_scheme_K4}
\end{align}
Finally, including radiation up to $G^3$, the second line of our formula \eqref{eq:impulse} agrees with the corresponding eikonal radiation reaction contribution in the spinless case \cite{Cristofoli:2021jas,DiVecchia:2022piu}, although our subtraction scheme for the exponentiation \eqref{eq:HEFT_exp} is different (see \cite{DiVecchia:2023frv} for a detailed comparison).
\vspace{-14pt}
\subsection{Waveform and spinning fluxes}

Using the mode expansion \eqref{eq:field_mode} for the graviton field, we now discuss the perturbative expansion of the waveform and the related fluxes (radiated linear and angular momentum) using the master equation \eqref{eq:master_formula}. Starting with the former, we observe that
\begin{align}
   &\hspace{-12pt}\langle \kappa h_{\mu \nu}(x) \rangle = \frac{\kappa \hbar^{3/2}}{4 \pi |\vec{x}|} \sum_{\sigma} \int \frac{\mathrm{d} \omega}{2 \pi} \Big(e^{i \omega u} \varepsilon^{\sigma}_{\mu \nu}(\hat{n}) \langle a_{\sigma}\left(\omega \hat{n}\right) \rangle \nonumber \\
   & \qquad  \qquad \qquad  \qquad  \qquad +e^{-i \omega u} \varepsilon^{*\sigma}_{\mu \nu}(\hat{n}) \langle a^{\dagger}_{\sigma}\left(\omega \hat{n}\right) \rangle \Big)
\end{align}
and therefore we need to evaluate only $ \langle a_{\sigma}\left(\omega \hat{n}\right) \rangle$ and its conjugate $ \langle a^{\dagger}_{\sigma}\left(\omega \hat{n}\right) \rangle$ using \eqref{eq:master_formula}. A straightforward calculation, valid up to 4PM order, gives
\begin{align}
    \label{eq:waveform}
   &\kappa \hbar^{3/2}\langle a_{\sigma}\left(\omega \hat{n}\right) \rangle \Big|_{\mathcal{O}(G^4)} \hspace{-10pt}= i \kappa \widetilde{\mathcal{K}}_{5, \mathcal{R}}^{\mathrm{cl}}\left(\omega \hat{n}\right) \nonumber \\
   &+ i \kappa \sum_{n=2}^{+\infty} \frac{1}{n!} \big\{ (\widetilde{\mathcal{K}}^{\mathrm{cl}})^{\odot (n-1)}, \widetilde{\mathcal{K}}_{5, \mathcal{R}}^{\mathrm{cl}}\left(\omega \hat{n}\right) \big\}_{\DB} \nonumber \\
  & + \frac{\kappa}{2} \int_k  \widetilde{\mathcal{K}}_{5, \mathcal{R}}^{*\mathrm{cl}}\left(k\right) \widetilde{\mathcal{K}}_{6, \mathcal{R}}^{A\mathrm{cl}}\left(\omega \hat{n},k\right) \nonumber \\
  & -\frac{\kappa}{4} \int_k (\widetilde{\mathcal{K}}_{6, \mathcal{R}}^{B\mathrm{cl}}\left(\omega \hat{n},k\right) + \widetilde{\mathcal{K}}_{6, \mathcal{R}}^{B*\mathrm{cl}}\left(\omega \hat{n},k\right)) \widetilde{\mathcal{K}}_{5, \mathcal{R}}^{\mathrm{cl}}\left(k\right)\,, 
\end{align}
where we have used the orthogonality of states with different number of gravitons. To obtain $ \langle a^{\dagger}_{\sigma}\left(\omega \hat{n}\right) \rangle$, we just take the conjugate of \eqref{eq:waveform}.

While the leading contribution in \eqref{eq:waveform} corresponds to the five-point amplitude identified in \cite{Cristofoli:2021vyo}, higher order contributions naturally involve iterations of the brackets with the four-point kernel and also potential contributions from the six-point and higher-point kernels. We notice, in agreement with \cite{Kim:2025hpn}, that the one-loop waveform contribution can be fully reproduced by the first and second contribution in \eqref{eq:waveform}. Using
\begin{align}
\{\widetilde{\mathcal{K}}^{\mathrm{cl}},\cdot\}_{\DB} &= \{\widetilde{\mathcal{K}}^{\mathrm{cl}},b_j^{\mu}\}_{\DB} \frac{\partial}{\partial b_j^{\mu}} + \{\widetilde{\mathcal{K}}^{\mathrm{cl}},v_j^{\mu}\}_{\DB} \frac{\partial}{\partial v_j^{\mu}} \nonumber \\
&+ \{\widetilde{\mathcal{K}}^{\mathrm{cl}},s_j^{\mu}\}_{\DB} \frac{\partial}{\partial s_j^{\mu}} \,,
\end{align}
we find that the term $\{ \widetilde{\mathcal{K}}^{\mathrm{cl}},\widetilde{\mathcal{K}}_{5, \mathcal{R}}^{\mathrm{cl}}\left(\omega \hat{n}\right) \}_{\DB} $ in the spinless case corresponds to the rotation of the tree-level waveform identified in \cite{Georgoudis:2023eke,Bini:2024rsy,Georgoudis:2024pdz}. We expect this to be also relevant for spinning one-loop waveforms \cite{Bohnenblust:2023qmy,Bohnenblust:2025gir}. On a formal level, the Dirac brackets effectively implement the causality prescription, bypassing the subtleties involved in the evaluation of KMOC cut contributions \cite{Caron-Huot:2023vxl}. It would be interesting to understand whether and how the six-point and higher-point kernels enter in \eqref{eq:waveform} for the two-loop waveform computation.\footnote{ In particular, note that the term on the third line in \eqref{eq:waveform} involving the six-point kernel would naturally contribute, together with a genuine two-loop contribution to $\widetilde{\mathcal{K}}_{5, \mathcal{R}}^{\mathrm{cl}}$ generated by gluing a five-point and a six-point tree amplitude across a three-particle cut \cite{Alessio:2025xx}, to the non-linear memory effect \cite{Christodoulou:1991cr,Wiseman:1991ss,Thorne:1992sdb} for the waveform which starts at 3PM order \cite{Alessio:2024onn}. 
} 

We now focus on the spinning fluxes using the master formula \eqref{eq:master_formula}. Starting with the radiated momentum 
\begin{align}
\mathbb{K}^{\mu} = \sum_{\sigma} \int_k \, k^{\mu} \, a^{\dagger}_{\sigma}\left(k\right) a_{\sigma}\left(k\right)\,,
\end{align}
we obtain the expression valid up to 5PM order
\begin{align}
\label{eq:radiated_momentum}
    &\hspace{-10pt}\langle \mathbb{K}^{\mu} \rangle \Big|_{\mathcal{O}(G^5)} = \int_k k^{\mu} \widetilde{\mathcal{K}}_{5, \mathcal{R}}^{*\mathrm{cl}} \widetilde{\mathcal{K}}_{5, \mathcal{R}}^{\mathrm{cl}}   \\
    &\quad + \frac{1}{2} \int_k k^{\mu} \left[\widetilde{\mathcal{K}}_{5, \mathcal{R}}^{\mathrm{cl}} \{ \widetilde{\mathcal{K}}^{\mathrm{cl}},\widetilde{\mathcal{K}}_{5, \mathcal{R}}^{*\mathrm{cl}} \}_{\DB} +\widetilde{\mathcal{K}}_{5, \mathcal{R}}^{*\mathrm{cl}} \{ \widetilde{\mathcal{K}}^{\mathrm{cl}},\widetilde{\mathcal{K}}_{5, \mathcal{R}}^{\mathrm{cl}} \}_{\DB} \right] \nonumber \\
     &\quad + \frac{i}{2} \int_{k_1,k_2} k_1^{\mu} \Big[\widetilde{\mathcal{K}}_{5, \mathcal{R}}^{\mathrm{cl}}(k_1) \widetilde{\mathcal{K}}_{5, \mathcal{R}}^{\mathrm{cl}}(k_2) \widetilde{\mathcal{K}}_{6, \mathcal{R}}^{A*\mathrm{cl}}\left(k_1,k_2\right) \nonumber \\
     & \qquad \qquad  \qquad\quad- \widetilde{\mathcal{K}}_{5, \mathcal{R}}^{*\mathrm{cl}}(k_1) \widetilde{\mathcal{K}}_{5, \mathcal{R}}^{*\mathrm{cl}}(k_2) \widetilde{\mathcal{K}}_{6, \mathcal{R}}^{A\mathrm{cl}}\left(k_1,k_2\right) \Big]\,. \nonumber 
\end{align}
Note that we could have derived \eqref{eq:radiated_momentum} by using momentum conservation, given that $\langle \mathbb{K}^{\mu} \rangle + m_1 \langle \Delta v_1^{\mu} \rangle + m_2 \langle \Delta v_2^{\mu} \rangle =0$. Indeed, a quick computation reveals that the symmetry properties of the Dirac brackets under the exchange $1 \leftrightarrow 2$ imply the cancellation of many contributions in \eqref{eq:impulse}, leaving behind only \eqref{eq:radiated_momentum} (see appendix \ref{sec:app1} for details).

In a similar way, we can also compute the radiated angular momentum \cite{Manohar:2022dea,DiVecchia:2022owy} using the operator
\begin{align}
\label{eq:KJ-GR}
&\mathbb{J}^{\mu \nu} = \sum_{\sigma = \pm } \int_k \varepsilon^{\alpha \alpha'}_{\sigma}(k) a^{\dagger}_{\sigma}(k)\left[(\mathcal{J})^{\mu \nu}_{\alpha \alpha' \beta \beta'}
\right] \varepsilon^{*\beta \beta'}_{\sigma}(k) a_{\sigma}(k) \,, \nonumber \\
&\quad \mathcal{J}^{\mu \nu}_{\alpha \alpha' \beta \beta'} =- i\eta_{\alpha \beta} \eta_{\alpha' \beta'} k^{[\mu} \frac{\stackrel{\leftrightarrow}{\partial}}{\partial k_{\nu]}}  
-2 i \eta_{\alpha' \beta'} \delta^{[\mu}_{\alpha} \delta^{\nu]}_{\beta} \,,
\end{align}
where $f \overset\leftrightarrow{\partial} g = f \partial g - g \partial f$, which is independent of any SSC condition unlike the single particle expressions \eqref{eq:angular_momentum}. To verify angular momentum conservation, following \cite{Jakobsen:2022zsx}, we convert the angular momentum to the vector representation in the center-of-mass (CM) frame,
\begin{align}
\label{eq:spinvectCM}
    J^\mu := \frac{1}{2 E} \epsilon^{\mu}{}_{\nu \rho\sigma} J^{\nu\rho} (m_1 v_1 + m_2 v_2)^{\sigma},
\end{align}
with the CM total energy $E:= \sqrt{m_1^2 + m_2^2 + 2 m_1 m_2 \sigma}$. We verify that the angular momentum is conserved $\langle \mathbb{J}^{\mu} \rangle + \Delta J_1^\mu + \Delta J_2^\mu = 0$. Obviously, this formalism applies also to the differential fluxes in the same way; see \cite{Gonzo:2023cnv} for example for their covariant operator formulation.

\section{2PM spin kick and angular momentum to higher orders in spin}\label{sec:angular_momentum}

In this section, we now apply our equation \eqref{eq:impulse} for the calculation of the impulse, the spin kick and the angular momentum for generic spinning binaries at 2PM order. To do so, we need to consider both the 2MPI 4-point kernel $\mathcal{K}^{\mathrm{cl}}$ and the 5-point one $\mathcal{K}_{5, \mathcal{R}}^{\mathrm{cl}}$. First, we write
\begin{align}
\label{eq:K4_2PM}
&\widetilde{\mathcal{K}}^{\mathrm{cl}}(b)\Big|_{\mathcal{O}(G^2)} = \int  \hat{\mathrm{d}}^4 q \,\hat{\delta}\left(2 \bar{p}_1 \cdot q\right) \hat{\delta}\left(2 \bar{p}_2 \cdot q\right) \mathrm{e}^{i\left(q \cdot b\right) / \hbar} \nonumber \\
&\qquad \qquad \quad \times \left(\mathcalMtree + \mathcal{M}_{4,\bigtriangleup}^{(1)} + \mathcal{M}_{4,\bigtriangledown}^{(1)}\right)\,,
\end{align}
in terms of the classical tree-level t-channel diagram $\mathcalMtree$ and the one-loop amplitude coming from the triangle topologies $\mathcal{M}_{4,\bigtriangleup}^{(1)} + \mathcal{M}_{4,\bigtriangledown}^{(1)}$. Up to 2PM order, the eikonal and the conservative $\hat{N}$ matrix element are equivalent \cite{DiVecchia:2023frv}, and here we directly import the recent one-loop results at higher orders in spin \cite{Bohnenblust:2024hkw} (see also \cite{Bern:2020buy,Liu:2021zxr,Kosmopoulos:2021zoq,Aoude:2022thd,Bautista:2023szu,Ben-Shahar:2023djm,Aoude:2023vdk,Luna:2023uwd,Gatica:2023iws,Gatica:2024mur,Chen:2024mmm,Haddad:2024ebn,Bonocore:2024uxk,Bonocore:2025stf}). Our calculation relies on the same assumptions as in \cite{Cangemi:2023ysz, Cangemi:2023bpe, Bohnenblust:2024hkw} regarding the contact terms in the Compton amplitude beyond quartic order in spin, which stem from unresolved ambiguities in the matching with the corresponding wave scattering problem on Kerr black holes.

When accounting for radiation, it is useful to distinguish between contributions from positive-energy gravitons -- hereafter referred to as dynamical -- and those from zero-energy gravitons, which we call static. For example, the radiative kernel admits the decomposition \cite{DiVecchia:2022owy,DiVecchia:2022piu,Adamo:2024oxy}
\begin{align}
   &\hspace{-15pt}\mathcal{K}^{\mathrm{cl}}_{5, \mathcal{R}}\left(k\right) \! = \!\lim _{\omega^* \rightarrow 0} \Bigg[ \Theta\left(\omega^*-\omega\right) \mathcal{K}_{5, \mathcal{R}}^{\mathrm{stat}}(k) \nonumber \\
   &\qquad \qquad\qquad \,\,+\Theta\left(\omega-\omega^*\right) \mathcal{K}_{5, \mathcal{R}}^{\mathrm{dyn}}(k)\Bigg]\,,
   \label{eq:stat_dyn}
\end{align}
where $\omega^*$ denote the graviton frequency cutoff marking the separation between the two terms.\footnote{This decomposition can be better understood at the off-shell level \cite{Biswas:2024ept}, we choose to focus on the on-shell S-matrix in this work.} We note that only the static part contributes at 2PM order, and it can be written in terms of the Weinberg soft factor  \cite{Weinberg:1965nx}
\begin{align}
\label{eq:K5_2PM}
   \mathcal{K}_{5, \mathcal{R}}^{\mathrm{stat}}(k)  = i \frac{\kappa}{2} \sum_{j=1,2,1',2'} \frac{\varepsilon_{\mu \nu}(k) \, p_{j}^\mu p_{j}^\nu}{\eta_j p_{j} \cdot k-i 0}\,,
\end{align}
where $\eta_j=1$ if $j=1',2'$ and $\eta_j=-1$ if $j=1,2$, which is therefore sensitive only to the initial $p^{\mu}_1,p^{\mu}_2$ and final momenta $p_1^{\prime \mu}$, $p_2^{\prime \mu}$. In the conservative case we express the final momenta in terms of the initial one
\begin{align}
\label{eq:classmomcon}
p_j^{\prime \mu} = p^{\mu}_j + (-1)^{j} Q^{\mu}(b,v_1,v_2,s_1,s_2)\,,
\end{align}
where $Q^{\mu}$ is the conservative 2PM spinning impulse which depends only on $b^{\mu} = b_2^{\mu} - b_1^{\mu}$.

Using the results for the kernels \eqref{eq:K4_2PM} and \eqref{eq:K5_2PM}, we have then computed using \eqref{eq:impulse} the impulse and the spin kick up to $\mathcal{O}(G^2 s_1^{j_1} s_2^{j_2})$ with $j_1+ j_2 \leq 11$
\begin{align}
\Delta v_i^{\mu}(b,v_j,s_j) \Big|_{\mathcal{O}(G^2)} \,, \qquad \Delta s_i^{\mu}(b,v_j,s_j) \Big|_{\mathcal{O}(G^2)} \,.
\end{align}
We recover here the impulse computed in \cite{Bohnenblust:2024hkw}, while the spin kick is new and generalizes the quadratic in spin result in \cite{Jakobsen:2021zvh}.  Moreover, we have computed the variation of the single particle angular momentum in \eqref{eq:Ji_expansion} up to $\mathcal{O}(G^2 s_1^{j_1} s_2^{j_2})$ with $j_1+ j_2 \leq 11$ by combining the dynamical contribution derived from the radial action
\begin{align}
\hspace{-15pt}\langle \Delta J_i^{\mu \nu}  \rangle^{\mathrm{dyn}} \Big|_{\mathcal{O}(G^2)} \hspace{-5pt} =  \{\widetilde{\mathcal{K}}^{\mathrm{cl}}, J_i^{\mu \nu} \}_{\DB} {+} \frac{1}{2!} \{(\widetilde{\mathcal{K}}^{\mathrm{cl}})^{\odot2}, J_i^{\mu \nu} \}_{\DB} \,,
\label{eq:dyn_Ji_2PM}
\end{align}
with the static angular momentum coming from the radiative bracket contribution\footnote{We defined $\int_k^* := \lim_{\omega^* \rightarrow 0} \int\mathrm{d}^4k\,\hat{\delta}(k^2)\Theta(k^0)\Theta(\omega^*-\omega)$.}
\begin{align}
&\hspace{-10pt}\langle \Delta J_i^{\mu \nu}  \rangle^{\mathrm{stat}} \Big|_{\mathcal{O}(G^2)}  =   \frac{i}{2!} \int_k^* \Big(\widetilde{\mathcal{K}}_{5, \mathcal{R}}^{\mathrm{cl}}(k) \{ \widetilde{\mathcal{K}}_{5, \mathcal{R}}^{*\mathrm{cl}}(k), J_i^{\mu \nu}\}_{\DB}  \nonumber\\
    & \qquad \qquad \qquad \qquad - \widetilde{\mathcal{K}}_{5, \mathcal{R}}^{*\mathrm{cl}}(k) \{ \widetilde{\mathcal{K}}_{5, \mathcal{R}}^{\mathrm{cl}}(k), J_i^{\mu \nu}\}_{\DB} \Big) \,.
    \label{eq:stat_Ji_2PM}
\end{align}
We present an exact derivation of the static contribution at all orders in spin in appendix~\ref{sec:app2}, which gives
\begin{align}
\langle \Delta J_i^{\mu\nu}\rangle^{\mathrm{stat}}\Big|_{\mathcal{O}(G^2)}&=-2 p_i^{[\mu} \langle \Delta b_i^{\nu]}\rangle^{\mathrm{stat}}\Big|_{\mathcal{O}(G^2)} \nonumber \\
&= -\mathrm{sgn}_i \frac{\kappa^2}{32\pi}\mathcal{I}(\sigma)p_i^{[\mu}Q^{\nu]}\,.
\end{align}
where $\mathcal{I}(\sigma)$ is defined in \eqref{eq:Isigma_int}. Interestingly, our calculation reveals that the static angular momentum change stems from the change of the impact parameter $\Delta b_i^\mu$, implying that it is sensitive to the Bondi–Metzner–Sachs frame \cite{Bondi:1960jsa,Sachs:1962wk,Veneziano:2022zwh,Riva:2023xxm,Elkhidir:2024izo,Veneziano:2025ecv}, which might also be related to the subtraction scheme \eqref{eq:subtraction_scheme_K4} of the radial action.  In the spinless case our results are consistent with the ones obtained in the eikonal formalism \cite{DiVecchia:2022owy,DiVecchia:2022piu,Heissenberg:2024umh}, but their generalization to higher orders in spin is completely new.

The explicit expressions of our results are given in the ancillary files. Specifically, we provide the the velocity change $\Delta v_1^\mu$, spin kick $\Delta s_1^\mu$, as well as the change of the impact parameter $\Delta b_1^\mu$ (including radiative contributions), which are sufficient to reconstruct the orbital angular momentum change $\Delta L_1^{\mu\nu} = (b_1^{[\mu} + \Delta b_1^{[\mu}) (v_1^{\nu]} + \Delta v_1^{\nu]}) - b_1^{[\mu} v_1^{\nu]} $. For completeness, we also provide the change of the total angular momentum vector in the CM frame $\Delta L^\mu $ with $L^\mu = \epsilon^{\mu}{}_{\nu\rho\sigma} b^\nu v_1^\rho v_2^\sigma /E$.

\section{Conclusions and Outlook}\label{sec:conclusion}

The importance of being able to identify the minimum set of gauge-invariant data required for a full description of the classical two-body problem cannot be overstated, as this provides an important bridge for the comparison between the different perturbative approaches (PM, PN and GSF) to various observables. 

An important step forward in this endeavor is a formulation of what is the classical generating functional of all observables, starting from the scattering scenario where the S-matrix description is available.  Here, we have combined three powerful techniques to achieve this task: the KMOC formalism, the exponential representation of the S-matrix and the newly introduced Dirac brackets. 

First, inspired by the eikonal formalism \cite{DiVecchia:2023frv}, we have developed a coherent state expansion of the $\hat{N}$ operator. We introduced the classical $(4{+}N)$-point gauge-invariant kernels, making then contact with the conjectural exponentiation of the classical S-matrix in impact parameter space. This allows to obtain a novel representation of the classical S-matrix \eqref{eq:HEFT_new}, which can be combined with the KMOC formalism to write a master formula to compute all observables \eqref{eq:master_formula}, as in \cite{Damgaard:2023ttc}.

Exploiting the properties of the Hilbert space for the two-body kinematics, we proposed a method to turn commutators into Dirac brackets for the generic radiative case (see also \cite{Kim:2024svw,Kim:2025hpn} and our earlier work \cite{Gonzo:2024zxo}). The algebra of the graviton mode oscillators can then be resolved, yielding new formulae for the impulse, spin kick and angular momentum of each particle \eqref{eq:impulse} and for the waveform \eqref{eq:waveform} and related fluxes \eqref{eq:radiated_momentum}. We find that in principle these observables depend on the higher-point radiative kernels (with two or more gravitons), although such contribution is suppressed at the lowest order in which it appears \cite{Britto:2021pud,Damgaard:2021ipf}, consistently with the uncertainty principle \cite{Cristofoli:2021jas}. It is left for the future, however, 
to clarify this point at higher orders in perturbation theory.

Our results suggest that -- as in the eikonal formalism, but with the $\hat{N}$ operator prescription -- that in-in scattering observables can be derived from the gauge-invariant 4-point $\widetilde{\mathcal{K}}^{\mathrm{cl}}$ and $4+N$-point kernels $\widetilde{\mathcal{K}}_{4+N, \mathcal{R}}^{\mathrm{cl}}$ by applying the Dirac brackets. We tested our equations in various limits, computing also for the first time the 2PM spin kick and spinning angular momentum up to $\mathcal{O}(G^2 s_1^{j_1} s_2^{j_2})$ with $j_1+ j_2 \leq 11$ which are provided in ancillary file. Relatedly, we have also linked the radiated part of the angular momentum at 2PM to a non-zero variation of $\Delta b^{\mu}$ in the intrinsic frame, shedding light on its dependence on the BMS frame from a classical perspective.

Our work opens the door to many future investigations. As a first step, it would be nice to have a formal proof of our Dirac bracket formalism and to clarify the role of the higher-point radiative kernels. Second, our derivation suggests a generalization of the first law of black hole mechanics with dissipation and spin, where our observables are linked to the variation of our kernels in phase space. It would be nice to find an efficient way to obtain our kernels either using the PM, PN or GSF perturbative approaches; so far we showed only how the conservative kernel can be simply extracted from the impulse in appendix~\ref{sec:app4}. Finally, it is desirable to apply our Dirac bracket formalism at higher orders in perturbation theory (see also \cite{Akpinar:2025bkt} for a recent application).

\paragraph{Acknowledgments}
It is a pleasure to thank A. Georgoudis, C. Heissenberg, G. Mogull, A. Ochirov, R. Porto, A. Pound and G.Veneziano for insightful comments and discussions on the manuscript. We thank V. Del Duca, E.Rosi, I.Rothstein and M.Saavedra for collaboration on related topics \cite{Alessio:2025xx}. We also thank the authors of \cite{Bohnenblust:2024hkw} for sharing with us their results and for the comparison. The work of RG is supported by the Royal Society grant RF$\backslash$ERE$\backslash$231084. The work of CS is supported by the China Postdoctoral Science Foundation under Grant No. 2022TQ0346, and the National Natural Science Foundation of China under Grant No. 12347146. This research was supported in part by grant NSF PHY-2309135 to the Kavli Institute for Theoretical Physics (KITP).

\paragraph{Data availability}
The data that support the findings of this article are openly available \cite{data}.

\appendix
\section{Test of the Dirac brackets: momentum conservation}
\label{sec:app1}

In this appendix, we test momentum conservation for the impulse by showing the equivalence 
\begin{align}
    \langle \mathbb{K}^\mu \rangle  &= -\langle \Delta \bar{p}_1^{\mu} \rangle -  \langle \Delta \bar{p}_2^{\mu} \rangle  \,,
\end{align}
between \eqref{eq:impulse} and \eqref{eq:radiated_momentum}. First, note that given that $\widetilde{\mathcal{K}}^{\mathrm{cl}}$ depends only on $b^{\mu} = b_2^{\mu} - b_1^{\mu}$ we have
\begin{align}
    \{\widetilde{\mathcal{K}}^{\mathrm{cl}}(\sigma,b), \bar{p}_1^{\mu} + \bar{p}_2^{\mu}\}_{\DB} = 0\,,
\end{align}
because of the antisymmetric structure of the Dirac brackets \eqref{eq:Dirac_brackets_final}. Then, restricting for simplicity to 4PM order, we can use the linearity of the integration and of the brackets in \eqref{eq:impulse} to write 
\begin{align}
\label{eq:impulse_sum}
    &\langle \Delta \bar{p}_1^{\mu} + \Delta \bar{p}_2^{\mu} \rangle \Big|_{\mathcal{O}(G^4)}  \\
    &= \frac{i}{2!} \int_k \Big(\widetilde{\mathcal{K}}_{5, \mathcal{R}}^{\mathrm{cl}}(k) \{ \widetilde{\mathcal{K}}_{5, \mathcal{R}}^{*\mathrm{cl}}(k),  \bar{p}_1^{\mu} +  \bar{p}_2^{\mu}\}_{\DB} \nonumber\\
    & \qquad \qquad  - \widetilde{\mathcal{K}}_{5, \mathcal{R}}^{*\mathrm{cl}}(k) \{ \widetilde{\mathcal{K}}_{5, \mathcal{R}}^{\mathrm{cl}}(k),  \bar{p}_1^{\mu} +  \bar{p}_2^{\mu}\}_{\DB} \Big)  \nonumber \\
    & + \frac{2 i}{3!} \int_k \Big(\widetilde{\mathcal{K}}_{5, \mathcal{R}}^{\mathrm{cl}}(k) \{ \widetilde{\mathcal{K}}^{\mathrm{cl}},\{\widetilde{\mathcal{K}}_{5, \mathcal{R}}^{*\mathrm{cl}}(k) ,  \bar{p}_1^{\mu} +  \bar{p}_2^{\mu}\}_{\DB}\}_{\DB} \nonumber \\
    & \qquad \qquad  -\widetilde{\mathcal{K}}_{5, \mathcal{R}}^{*\mathrm{cl}}(k) \{\widetilde{\mathcal{K}}^{\mathrm{cl}} ,\{ \widetilde{\mathcal{K}}_{5, \mathcal{R}}^{\mathrm{cl}}(k),  \bar{p}_1^{\mu} +  \bar{p}_2^{\mu}\}_{\DB}\}_{\DB} \Big)  \nonumber \\
    & + \frac{i}{3!} \int_k \Big(\{ \widetilde{\mathcal{K}}^{\mathrm{cl}},\widetilde{\mathcal{K}}_{5, \mathcal{R}}^{\mathrm{cl}}(k) \}_{\DB} \{\widetilde{\mathcal{K}}_{5, \mathcal{R}}^{*\mathrm{cl}}(k) ,  \bar{p}_1^{\mu} +  \bar{p}_2^{\mu}\}_{\DB} \nonumber \\
    & \qquad \qquad -\{ \widetilde{\mathcal{K}}^{\mathrm{cl}},\widetilde{\mathcal{K}}_{5, \mathcal{R}}^{*\mathrm{cl}}(k) \}_{\DB} \{ \widetilde{\mathcal{K}}_{5, \mathcal{R}}^{\mathrm{cl}}(k),  \bar{p}_1^{\mu} +  \bar{p}_2^{\mu}\}_{\DB} \Big)\,. \nonumber
\end{align}
We now use the definition of $\widetilde{\mathcal{K}}_{5, \mathcal{R}}^{\mathrm{cl}} $ in \eqref{eq:Fourier_transform}
\begin{align}
\label{eq:fourier_trasformK5}
&\hspace{-5pt}\widetilde{\mathcal{K}}_{5, \mathcal{R}}^{\mathrm{cl}}\left(b_1, b_2;k_1\right) \!=\! \int \hat{\mathrm{d}}^4 q_1 \hat{\mathrm{d}}^4 q_2 \,  \hat{\delta}\left(2 \bar{p}_1 \cdot q_1\right) \hat{\delta}\left(2 \bar{p}_2 \cdot q_2\right)  \\
& \qquad \times \hat{\delta}^4(q_1+q_2-k_1)\mathrm{e}^{i\left(q_1 \cdot b_1+q_2 \cdot b_2\right) / \hbar} \mathcal{K}_{5, \mathcal{R}}^{\mathrm{cl}}\left(q_1, q_2\right)\,,  \nonumber
\end{align}
to show how we can simplify the action of the Dirac brackets \eqref{eq:Dirac_brackets_final} at the integrand level. Indeed, because the integrand representation of \eqref{eq:fourier_trasformK5} still contains the transversality constraints, the action of the Dirac brackets reduce to the Poisson brackets $\{ \cdot, \cdot\}_{\DB} \to \{ \cdot, \cdot\}_{\PB} $
\begin{align}
&\{ \widetilde{\mathcal{K}}_{5, \mathcal{R}}^{\mathrm{cl}}, \bar{p}_1^{\mu} + \bar{p}_2^{\mu}\}_{\DB}  \\
& = \int \hat{\mathrm{d}}^4 q_1 \hat{\mathrm{d}}^4 q_2 \,\hat{\delta}^4(q_1+q_2-k)  \hat{\delta}\left(2 \bar{p}_1 \cdot q_1\right) \hat{\delta}\left(2 \bar{p}_2 \cdot q_2\right) \nonumber \\
& \qquad \times  \{ \mathrm{e}^{i\left(q_1 \cdot b_1+q_2 \cdot b_2\right) / \hbar} , \bar{p}_1^{\mu} + \bar{p}_2^{\mu} \}_{\PB}  \mathcal{K}_{5, \mathcal{R}}^{\mathrm{cl}}\left(q_1, q_2\right)  \nonumber \\
&= \int \hat{\mathrm{d}}^4 q_1 \hat{\mathrm{d}}^4 q_2 \,\hat{\delta}^4(q_1+q_2-k_1)  \hat{\delta}\left(2 \bar{p}_1 \cdot q_1\right) \hat{\delta}\left(2 \bar{p}_2 \cdot q_2\right) \nonumber \\
& \qquad \times \mathrm{e}^{i\left(q_1 \cdot b_1+q_2 \cdot b_2\right) / \hbar}  (i q_1^{\mu}+i q_2^{\mu}) \mathcal{K}_{5, \mathcal{R}}^{\mathrm{cl}}\left(q_1, q_2\right)  \nonumber \\
&= \int \hat{\mathrm{d}}^4 q_1 \hat{\mathrm{d}}^4 q_2 \,\hat{\delta}^4(q_1+q_2-k_1)  \hat{\delta}\left(2 \bar{p}_1 \cdot q_1\right) \hat{\delta}\left(2 \bar{p}_2 \cdot q_2\right) \nonumber \\
& \qquad \times  \mathrm{e}^{i\left(q_1 \cdot b_1+q_2 \cdot b_2\right) / \hbar}  (i k^{\mu}) \mathcal{K}_{5, \mathcal{R}}^{\mathrm{cl}}\left(q_1, q_2\right)   \nonumber \\
&= i k^{\mu}  \widetilde{\mathcal{K}}_{5, \mathcal{R}}^{\mathrm{cl}}\,.
\end{align}
Taking the conjugate of this expression we also obtain $\{ \widetilde{\mathcal{K}}_{5, \mathcal{R}}^{*\mathrm{cl}}, \bar{p}_1^{\mu} + \bar{p}_2^{\mu}\}_{\DB}  = -i k^{\mu}  \widetilde{\mathcal{K}}_{5, \mathcal{R}}^{*\mathrm{cl}} $, which allows us to combine the remaining contributions in \eqref{eq:impulse_sum} as
\begin{align}
\label{eq:impulse_sum2}
    &\langle  \Delta \bar{p}_1^{\mu} + \Delta \bar{p}_2^{\mu} \rangle \Big|_{\mathcal{O}(G^4)} = -\int_k k^{\mu} \widetilde{\mathcal{K}}_{5, \mathcal{R}}^{*\mathrm{cl}} \widetilde{\mathcal{K}}_{5, \mathcal{R}}^{\mathrm{cl}}   \\
    &\quad - \frac{1}{2} \int_k k^{\mu} \left[\widetilde{\mathcal{K}}_{5, \mathcal{R}}^{\mathrm{cl}} \{ \widetilde{\mathcal{K}}^{\mathrm{cl}},\widetilde{\mathcal{K}}_{5, \mathcal{R}}^{*\mathrm{cl}} \}_{\DB} +\widetilde{\mathcal{K}}_{5, \mathcal{R}}^{*\mathrm{cl}} \{ \widetilde{\mathcal{K}}^{\mathrm{cl}},\widetilde{\mathcal{K}}_{5, \mathcal{R}}^{\mathrm{cl}} \}_{\DB} \right]\,, \nonumber
\end{align}
giving back \eqref{eq:radiated_momentum} as desired. Following the same steps, we can generalize the arguments made here for the canonical angular momentum (c.f. \cite{Jakobsen:2022zsx}) to obtain
\begin{align}
\langle \mathbb{J}^{\mu \nu} \rangle = -\langle \Delta J_{1,\mathrm{can}}^{\mu \nu} \rangle - \langle \Delta J_{2,\mathrm{can}}^{\mu \nu}\rangle\,.
\end{align}

\section{Details about the static angular momentum integrals}
\label{sec:app2}

In this appendix, we evaluate the static contribution to the variation of the angular momentum at 2PM order and at all orders in spin using \eqref{eq:stat_Ji_2PM} and the radiative kernel \eqref{eq:K5_2PM}, which is expressed in terms of the incoming kinematics and the conservative impulse \cite{Guevara:2018wpp}
\begin{align}
\nonumber Q^{\mu}(b,v_j,s_j) &= -\frac{G m_1 m_2}{\sqrt{\sigma^2 -1 }} \sum_{h=\pm}f_h(\sigma)\frac{b_h^{\mu}}{b_h^2}\,, \\\nonumber
\qquad f_h(\sigma)&=(2\sigma^2 -1+ 2 h \sigma \sqrt{\sigma^2 - 1})\,,
\nonumber \\
\qquad \qquad b_{\pm}^{\mu}&=\Pi^{\mu \nu} b_{\nu}\pm \frac{\epsilon^\mu_{\,\,\,\nu\rho\lambda} s_+^{\nu} v_1^{\rho} v_2^{\lambda}}{\sqrt{\sigma^2-1}}\,,
\label{eq:2PM_impulse}
\end{align}
where $s^{\mu}_+ = s^{\mu}_1+ s^{\mu}_2 $ and $\Pi^{\mu \nu}$ is the two-body projector
\begin{align}
\label{eq:twobody_proj}
     \Pi^{\mu \nu} = \eta^{\mu \nu} - \frac{1}{\sigma^2 - 1}\left(v^{\mu}_{1} v^{\nu}_{1} + v^{\mu}_{2} v^{\nu}_{2} - 2\sigma v^{(\mu}_{1} v^{\nu)}_{2}\right)\,.
\end{align}
The advantage of writing \eqref{eq:2PM_impulse} in terms of the projector is that there is a simplification happening when we consider the Dirac brackets of two functions $J_1^{\mu \nu}$ and $\widetilde{\mathcal{K}}_{5, \mathcal{R}}^{*\mathrm{cl}}$ which are invariant under 
\begin{align}
    b_1^\mu \to b_1^\mu + \alpha_1 v_1^{\mu}\,, \qquad b_2^\mu \to b_2^\mu + \alpha_2 v_2^{\mu}\,,
\end{align}
namely that their Dirac bracket \eqref{eq:Dirac_brackets_final} reduces to the primed ones \eqref{eq:primed_brackets} (see appendix \ref{sec:app3} for a full proof)
\begin{align}
     \{\widetilde{\mathcal{K}}_{5, \mathcal{R}}^{*\mathrm{cl}}, J_1^{\mu \nu}\}_{\DB} \to \{\widetilde{\mathcal{K}}_{5, \mathcal{R}}^{*\mathrm{cl}}, J_1^{\mu \nu}\}^{\prime} \,. 
\end{align}
Hence, we can simplify the Dirac bracket in \eqref{eq:stat_Ji_2PM} to
 \begin{align}
    \{\widetilde{\mathcal{K}}_{5, \mathcal{R}}^{*\mathrm{cl}}, J_1^{\mu \nu}\}_{\DB} &= \{\widetilde{\mathcal{K}}_{5, \mathcal{R}}^{*\mathrm{cl}}, J_1^{\mu \nu}\}^{\prime}  \\
    &\hspace{-40pt}= \frac{\partial \widetilde{\mathcal{K}}_{5, \mathcal{R}}^{*\mathrm{cl}}}{\partial b_1^{\alpha}}\frac{\partial J_1^{\mu \nu}}{\partial b_1^{\beta}} \{b_1^{\alpha}, b_1^{\beta}\}'
    + \frac{\partial \widetilde{\mathcal{K}}_{5, \mathcal{R}}^{*\mathrm{cl}}}{\partial v_1^{\alpha}} \frac{\partial J_1^{\mu \nu}}{\partial b_1^{\beta}} \{v_1^{\alpha}, b_1^{\beta}\}' \nonumber  \\
    &\hspace{-40pt}+ \frac{\partial \widetilde{\mathcal{K}}_{5, \mathcal{R}}^{*\mathrm{cl}}}{\partial s_1^{\alpha}} \frac{\partial J_1^{\mu \nu}}{\partial b_1^{\beta}} \{s_1^{\alpha}, b_1^{\beta}\}'
    + \frac{\partial \widetilde{\mathcal{K}}_{5, \mathcal{R}}^{*\mathrm{cl}}}{\partial b_1^{\alpha}} \frac{\partial J_1^{\mu \nu}}{\partial v_1^{\beta}} \{b_1^{\alpha}, v_1^{\beta}\}'   \nonumber  \\
    &\hspace{-40pt}+ \frac{\partial \widetilde{\mathcal{K}}_{5, \mathcal{R}}^{*\mathrm{cl}}}{\partial b_1^{\alpha}}\frac{\partial J_1^{\mu \nu}}{\partial s_1^{\beta}}  \{b_1^{\alpha}, s_1^{\beta}\}'
    + \frac{\partial \widetilde{\mathcal{K}}_{5, \mathcal{R}}^{*\mathrm{cl}}}{\partial s_1^{\alpha}} \frac{\partial J_1^{\mu \nu}}{\partial s_1^{\beta}} \{s_1^{\alpha},s_1^{\beta}\}' \,. \nonumber
\end{align}
Evaluated on the static contribution, this becomes
 \begin{align}
    &\hspace{-20pt}\{\widetilde{\mathcal{K}}_{5, \mathcal{R}}^{*\mathrm{cl}}, J_1^{\mu \nu}\}_{\DB}
    = \\
    & 2 b_1^{[\mu} \frac{\partial \widetilde{\mathcal{K}}^{*\mathrm{stat}}_{5,\mathcal{R}}}{\partial b_1^{\nu]}}
      + 2 v_1^{[\mu} \frac{\partial \widetilde{\mathcal{K}}^{*\mathrm{stat}}_{5,\mathcal{R}}}{\partial v_1^{\nu]}}
      + 2 s_1^{[\mu} \frac{\partial \widetilde{\mathcal{K}}^{*\mathrm{stat}}_{5,\mathcal{R}}}{\partial s_1^{\nu]}} \nonumber
\end{align}
We can then evaluate \eqref{eq:stat_Ji_2PM} explicitly
\begin{align}
    &\langle \Delta J_1^{\mu \nu}  \rangle^{\mathrm{stat}} \Big|_{\mathcal{O}(G^2)} =
    i \int_k^* \Bigg[
    b_1^{[\mu} \widetilde{\mathcal{K}}^{\mathrm{stat}}_{5,\mathcal{R}} \frac{\overset\leftrightarrow{\partial}}{\partial b_1^{\nu]}} \widetilde{\mathcal{K}}^{*\mathrm{stat}}_{5,\mathcal{R}}   \\
    &\qquad + v_1^{[\mu}  \widetilde{\mathcal{K}}^{\mathrm{stat}}_{5,\mathcal{R}} \frac{\overset\leftrightarrow{\partial}}{\partial v_1^{\nu]}} \widetilde{\mathcal{K}}^{*\mathrm{stat}}_{5,\mathcal{R}}
    + s_1^{[\mu}  \widetilde{\mathcal{K}}^{\mathrm{stat}}_{5,\mathcal{R}} \frac{\overset\leftrightarrow{\partial}}{\partial s_1^{\nu]}} \widetilde{\mathcal{K}}^{*\mathrm{stat}}_{5,\mathcal{R}} \nonumber
    \Bigg].
\end{align}
In the spinless limit, this expression resembles the proposal in eq.(6.1) of \cite{Heissenberg:2024umh}\footnote{Eq. (6.1) in \cite{Heissenberg:2024umh} is written in terms of $\tilde{u}_i$, which are classically different from $v_i$ we are using here.}. From the above equation we see that we need to compute the following integrals
\begin{align}
\label{eq:familyintegrals}-i\int_k^* \widetilde{\mathcal{K}}^{\mathrm{stat}}_{5,\mathcal{R}} \frac{\overset\leftrightarrow{\partial}}{\partial x_i^{\mu}}\widetilde{\mathcal{K}}^{*\mathrm{stat}}_{5,\mathcal{R}},\qquad x_i^{\mu}=(v_i^{\mu},s_i^{\mu},b_i^{\mu})
\end{align}
where the static kernel $\widetilde{\mathcal{K}}^{\mathrm{stat}}_{5,\mathcal{R}}$ is given in \eqref{eq:K5_2PM} and, since it depends on the initial and final momenta in \eqref{eq:classmomcon}, it has an implicit dependence on $(s_i,b_i)$ through the conservative impulse $Q^{\mu}(x_i)$ in \eqref{eq:2PM_impulse} and both an explicit and implicit dependence on $(v_i)$. Therefore
\begin{align}
\label{chainrules}
&\frac{\partial}{\partial v_1^{\mu}}\widetilde{\mathcal{K}}^{*\mathrm{stat}}_{5,\mathcal{R}}=m_1 \left( \frac{\partial}{\partial p_1^{\mu}}\widetilde{\mathcal{K}}^{*\mathrm{stat}}_{5,\mathcal{R}}  + \frac{\partial}{\partial p_1^{\prime \mu}}\widetilde{\mathcal{K}}^{*\mathrm{stat}}_{5,\mathcal{R}} \right) \nonumber \\
&\qquad\qquad\quad+m_1\frac{\partial Q^{\nu}}{\partial p_1^{\mu}}\frac{\partial}{\partial Q^{\nu}}\widetilde{\mathcal{K}}^{*\mathrm{stat}}_{5,\mathcal{R}},\\&
\frac{\partial}{\partial b_1^{\mu}}\widetilde{\mathcal{K}}^{*\mathrm{stat}}_{5,\mathcal{R}}=\frac{\partial Q^{\nu}}{\partial b_1^{\mu}}\frac{\partial}{\partial Q^{\nu}}\widetilde{\mathcal{K}}^{*\mathrm{stat}}_{5,\mathcal{R}},\\&
\frac{\partial}{\partial s_1^{\mu}}\widetilde{\mathcal{K}}^{*\mathrm{stat}}_{5,\mathcal{R}}=\frac{\partial Q^{\nu}}{\partial s_1^{\mu}}\frac{\partial}{\partial Q^{\nu}}\widetilde{\mathcal{K}}^{*\mathrm{stat}}_{5,\mathcal{R}},
\end{align}
with
\begin{align}
\frac{\partial}{\partial Q^{\nu}}\widetilde{\mathcal{K}}^{*\mathrm{stat}}_{5,\mathcal{R}}=\bigg(\frac{\partial}{\partial p_2^{\prime \nu}}-\frac{\partial}{\partial p_1^{\prime \nu}}\bigg)\widetilde{\mathcal{K}}^{*\mathrm{stat}}_{5,\mathcal{R}},
\end{align}
where we used \eqref{eq:classmomcon}. Hence, the family of integrals we need to compute are
\begin{align}
\mathcal{I}_{\mu,i}:= -i\int_k^* \widetilde{\mathcal{K}}^{\mathrm{stat}}_{5,\mathcal{R}} \frac{\overset\leftrightarrow{\partial}}{\partial p_i^{\mu}}\widetilde{\mathcal{K}}^{*\mathrm{stat}}_{5,\mathcal{R}},
\end{align}
giving (see e.g. \cite{DiVecchia:2023frv} for more details)
\begin{align}
\mathcal{I}^{\mu}_i=\frac{\kappa^2}{4}\sum_{j=1,2,1',2'}\alpha_{j,i}p_j^{\mu}+\beta_{j,i}p_i^{\mu},
\end{align}
with coefficients 
\begin{align}
    &\hspace{-15pt}\alpha_{j,i}=\pi\frac{2\eta_i(2\sigma^2_{ij}-1)+\sigma_{ij}\Delta_{ij}(\eta_j-\eta_i)(2\sigma^2_{ij}-3)}{4(2\pi)^2(\sigma^2_{ij}-1)},\\
    &\hspace{-15pt}\beta_{j,i}=\pi\frac{m_j}{m_i}\frac{\Delta_{ij}(\eta_j-\eta_i)-2\eta_j\sigma_{ij}(2\sigma^2_{ij}-1)}{4(2\pi)^2(\sigma^2_{ji}-1)},
\end{align}
and
\begin{align}
\Delta_{ij}=\frac{\arccosh{\sigma_{ij}}}{(\sigma^2_{ij}-1)^{\frac{1}{2}}},\qquad \sigma_{ij}=-v_i\cdot v_j,
\end{align}
with $\sigma_{12}=\sigma_{1'2'}=\sigma$, $\sigma_{ii}=\sigma_{i'i'}=1$, and
\begin{align}
\label{eq:PMsigma}
&\sigma_{12'}=\sigma_{21'}=\sigma-\frac{Q^2}{2 m_1m_2},\\&
\sigma_{11'}=1+\frac{Q^2}{2m_1^2},\qquad
\sigma_{22'}=1+\frac{Q^2}{2m_2^2},
\end{align}
as follows from \eqref{eq:classmomcon}. It is convenient to introduce
\begin{align}
\frac{1}{2}\mathcal{I}(\sigma) := \frac{8-5\sigma^2}{3(\sigma^2-1)}+\frac{\sigma(2\sigma^2-3)}{(\sigma^2-1)^{\frac{3}{2}}}\arccosh\sigma.
\label{eq:Isigma_int}
\end{align}
in terms of which we get\footnote{Notice that there is a discrepancy between our result in \eqref{eq:DeltaLspinless} and the one in \cite{DiVecchia:2022piu,DiVecchia:2023frv}. However, at the level of the four-dimensional vector in \eqref{eq:spinvectCM} they agree and therefore, in the CM frame, they are equal up to a $0i$ component.}
\begin{align}
\label{eq:DeltaLspinless}
&\langle \Delta L_1^{\mu \nu}  \rangle^{\mathrm{stat}} \Big|_{\mathcal{O}(G^2)}\nonumber \\
&\qquad=-\frac{\kappa^2}{16\pi}\frac{Gm_1m_2(2\sigma^2-1)}{b^2\sqrt{\sigma^2-1}}\mathcal{I}(\sigma)p_1^{[\mu}b^{\nu]},
\end{align}
in the spinless case, whereas in the spinning case we find
\begin{align}
\label{eq:DeltaLspin}
&\langle \Delta J_1^{\mu \nu}  \rangle^{\mathrm{stat}} \Big|_{\mathcal{O}(G^2)}\nonumber \\
&\qquad=-\frac{\kappa^2}{32\pi}\frac{G m_1m_2}{\sqrt{\sigma^2-1}}\mathcal{I}(\sigma)\sum_{h=\pm}f_h(\sigma)\frac{p_1^{[\mu}b_h^{\nu]}}{b^2_h},
\end{align}
which correctly reduces to \eqref{eq:DeltaLspinless} for $s_+=0$.
Both equations \eqref{eq:DeltaLspinless} and \eqref{eq:DeltaLspin} can be compactly written as
\begin{align}
\langle \Delta J_1^{\mu\nu}\rangle^{\mathrm{stat}}\Big|_{\mathcal{O}(G^2)}=\frac{\kappa^2}{32\pi}\mathcal{I}(\sigma)p_1^{[\mu}Q^{\nu]}.
\end{align}
This stems from the fact that the radiated impulse and spin kick vanish at 2PM, and the radiative contribution to angular momentum change is thus purely from
\begin{align}
    \langle \Delta b_1^{\mu}\rangle^{\mathrm{stat}} \Big|_{\mathcal{O}(G^2)} = -\frac{\kappa^2}{32\pi}\mathcal{I}(\sigma)Q^{\mu}.
\end{align}
The expression for particle 2 is obtained by simply replacing $1\leftrightarrow 2$ in the above equation and multiplying it by an overall minus sign. Then it is immediate to show that, going back to canonical spin variables, 
\begin{align}
\hspace{-12pt}\langle \mathbb{J}^{\mu\nu}\rangle\Big|_{\mathcal{O}(G^2)}= - \langle \Delta J_{1,\mathrm{can}}^{\mu\nu}\rangle\Big|_{\mathcal{O}(G^2)} -  \langle \Delta J_{2,\mathrm{can}}^{\mu\nu}\rangle\Big|_{\mathcal{O}(G^2)} \,,
\end{align}
where $\langle \mathbb{J}^{\mu\nu}\rangle\Big|_{\mathcal{O}(G^2)}$ was first computed in \cite{Alessio:2022kwv} (see \cite{Damour:2020tta,Mougiakakos:2021ckm,Jakobsen:2021smu,Gralla:2021qaf,DiVecchia:2022owy,DiVecchia:2022piu,Heissenberg:2024umh} for the spinless result, and \cite{Mogull:2025cfn} for an alternative derivation from the worldline trajectory ).

\section{From Dirac brackets to primed brackets}
\label{sec:app3}

We consider the generic Dirac brackets between two functions $\tilde{f}(b_j,v_j,s_j)$ and $\tilde{g}(b_j,v_j,s_j)$. We claim that if $\tilde{f}$, $\tilde{g}$ are invariant under
\begin{align}
    b_1^\mu \to b_1^\mu + \alpha_1 v_1^{\mu}\,, \qquad b_2^\mu \to b_2^\mu + \alpha_2 v_2^{\mu}\,,
    \label{eq:shift}
\end{align}
then the constrained Dirac brackets \eqref{eq:Dirac_brackets_final} reduce to the primed ones \eqref{eq:primed_brackets}.

First, we assume that $\tilde{f}$ is invariant under $b_1^\mu \to b_1^\mu + \alpha_1 v_1^{\mu}$. Infinitesimally, this means
\begin{align}
    0 &= \tilde{f}(b_1 + \alpha_1 v_1, \ldots) - \tilde{f}(b_1, \ldots) \nonumber \\
    &= \alpha_1 v_1^{\mu} \frac{\partial \tilde{f}(b_1,\ldots)}{\partial b_1^{\mu}} + \mathcal{O}(\alpha_1^2)\,, 
\end{align}
and repeating the argument for both shifts in \eqref{eq:shift} we obtain
\begin{align}
    v_1^{\mu} \frac{\partial \tilde{f}(b_1,b_2,\ldots)}{\partial b_1^{\mu}} =  v_2^{\mu} \frac{\partial \tilde{f}(b_1,b_2,\ldots)}{\partial b_2^{\mu}} = 0\,.
    \label{eq:simpl}
\end{align}
Clearly, the same holds for $\tilde{g}$ if it is invariant under \eqref{eq:shift}. This implies some simplification of the Dirac brackets
\begin{align}
 &\{\tilde{g},\tilde{f}\}_{\DB} =\!\! \sum_{i,j=1,2} \Big[ \frac{\partial \tilde{g}}{\partial b_i^{\alpha}} \frac{\partial \tilde{f}}{\partial b_j^{\beta}} \{b_i^{\alpha},b_j^{\beta}\}_{\DB}  + \frac{\partial \tilde{g}}{\partial b_i^{\alpha}} \frac{\partial \tilde{f}}{\partial v_j^{\beta}} \{b_i^{\alpha},v_j^{\beta}\}_{\DB}  \nonumber \\
 &\qquad\qquad+ \frac{\partial \tilde{g}}{\partial b_i^{\alpha}} \frac{\partial \tilde{f}}{\partial s_j^{\beta}} \{b_i^{\alpha},s_j^{\beta}\}_{\DB}  +  \frac{\partial \tilde{g}}{\partial v_i^{\alpha}} \frac{\partial \tilde{f}}{\partial b_j^{\beta}} \{v_i^{\alpha},b_j^{\beta}\}_{\DB}  \nonumber  \\
    &\qquad\qquad +  \frac{\partial \tilde{g}}{\partial s_i^{\alpha}}  \frac{\partial \tilde{f}}{\partial b_j^{\beta}} \{s_i^{\alpha},b_j^{\beta}\}_{\DB} +  \frac{\partial \tilde{g}}{\partial s_i^{\alpha}} \frac{\partial \tilde{f}}{\partial s_j^{\beta}} \{s_i^{\alpha},s_j^{\beta}\}_{\DB} \Big]\,,  \nonumber 
\end{align}
because \eqref{eq:simpl} sets to the zero some of the contributions proportional to $v_1^{\mu}$ and $v_2^{\mu}$ in the Dirac brackets \eqref{eq:Dirac_brackets_final}. Upon inspection, we find the brackets exactly reduce the primed brackets in \eqref{eq:primed_brackets}.

Finally, notice that if $\tilde{f}$ or $\tilde{g}$ depend only on the transverse components of $b^{\mu}=b_2^{\mu}-b_1^{\mu}$ (i.e., $b \cdot v_1 = b \cdot v_2 = 0$), we can write $b^{\mu}$ in terms of the projected vector $\Pi^{\mu \nu} b_{\nu}$, with the two-body projector $\Pi^{\mu \nu}$ in \eqref{eq:twobody_proj}. Doing so, we can make manifest the invariance under \eqref{eq:shift}, allowing to simplify the Dirac brackets to the primed ones.

\section{Extracting the radial action from the conservative impulse}
\label{sec:app4}
With the conservative formula \eqref{eq:impulse}, we can extract the $n$PM radial action directly from the $n$PM conservative momentum impulse.
Firstly, we subtract all the contributions from iterations of lower order radial action from the $n$PM impulse, obtaining the Dirac brackets of the $n$PM radial action and the initial momentum,
\begin{align}
    \label{eq_nPMDB}
    &\{\widetilde{\mathcal{K}}^{\mathrm{cl}}{}^{(n)}, \bar{p}_{1}^\mu \}_{\DB} = \Delta \bar{p}_1^{(n),\mu} \nonumber \\
    &\qquad \quad {-} \sum_{j=2}^{n} \frac{1}{j!}  \underbrace{ \{\widetilde{\mathcal{K}}^{\mathrm{cl}}, ...,  \{\widetilde{\mathcal{K}}^{\mathrm{cl}}}_{j \text{ times}}, \bar{p}_1^\mu\} ... \}\, \Big|_{n\mathrm{PM}} \,.
\end{align}
Similar to the spinless case, we essentially integrate \eqref{eq_nPMDB} to get $\widetilde{\mathcal{K}}^{\mathrm{cl}}{}^{(n)}$. We note that the radial action depends on the initial vectors $(b^\mu, v_1^\mu, v_2^\mu, s_1^\mu, s_2^\mu)$. We find it convenient to write the radial action in terms of the module $|b|$ and direction $\hat{b}^\mu$ of the impact parameter separately with $|b| := \sqrt{b^\mu b_\mu},\, \hat{b}^\mu:=b^\mu / |b|$. On our constraints, the Dirac bracket with $v_1^\mu$ can be written as,
\begin{align}
    \{\widetilde{\mathcal{K}}^{\mathrm{cl}}{}^{(n)}, \bar{p}_{1}^\mu \}_{\DB} =& \frac{\partial \widetilde{\mathcal{K}}^{\mathrm{cl}}{}^{(n)}}{\partial |b|} \{|b|, \bar{p}_{1}^\mu \}_{\DB}
    + \frac{\partial \widetilde{\mathcal{K}}^{\mathrm{cl}}{}^{(n)}}{\partial \hat{b}^\nu} \{\hat{b}^\nu, \bar{p}_{1}^\mu \}_{\DB} \nonumber\\
    =&\, \frac{\partial \widetilde{\mathcal{K}}^{\mathrm{cl}}{}^{(n)}}{\partial |b|} \hat{b}^\mu 
    + \frac{\partial \widetilde{\mathcal{K}}^{\mathrm{cl}}{}^{(n)}}{\partial \hat{b}^\nu} \frac{\hat{\varepsilon}^\nu \hat{\varepsilon}^\mu}{|b|}, &
\end{align}
where $\hat{\varepsilon}^\mu := \epsilon^{\mu}{}_{\nu\rho\sigma} \hat{b}^\nu v_1^\rho v_2^\sigma/ \sqrt{\sigma^2 -1}$ is a unit vector orthogonal to $b^\nu, v_1^\rho, v_2^\sigma$. We note that the term proportional to $\hat{b}^\mu$ comes only from the scalar derivative with respect to $|b|$. Taking of this advantage, we can just take the coefficient of $\hat{b}^\mu$, write it in terms of $(|b|, \hat{b}^\mu, v_1^\mu, v_2^\mu, s_1^\mu, s_2^\mu)$, and perform the integration over $|b|$ to obtain the radial action. Practically, this is trivial to do, as the radial action at a given PM and spin order is homogeneous in $|b|$ by dimensional analysis.

\bibliographystyle{apsrev4-1_title}
\bibliography{references}

\end{document}